\documentclass[epj]{svjour}
\usepackage{graphics}
\usepackage{url}
\usepackage{color}

\begin{document}
\title{Extracting Nuclear Symmetry Energies at High Densities \\from Observations of Neutron Stars and Gravitational Waves}
\author{Nai-Bo Zhang\inst{1,2} and Bao-An Li\inst{2}
}                     
\mail{Bao-An.Li@Tamuc.edu}          
\institute{Shandong Provincial Key Laboratory of Optical Astronomy and Solar-Terrestrial Environment, School of Space Science and Physics,
Institute of Space Sciences, Shandong University, Weihai, 264209, China \and Department of Physics and Astronomy, Texas A$\&$M University-Commerce, Commerce, TX 75429, USA}
\date{Received: date / Revised version: date}
%
\abstract{By numerically inverting the Tolman-Oppenheimer-Volkov (TOV) equation using an explicitly isospin-dependent parametric Equation of State (EOS) of dense neutron-rich nucleonic matter, a restricted EOS parameter space  is established using observational constraints on the radius, maximum mass, tidal deformability and causality condition of neutron stars (NSs). The constraining band obtained for the pressure as a function of energy (baryon) density is in good agreement with that extracted recently by the LIGO+Virgo Collaborations from their improved analyses of the NS tidal deformability in GW170817. Rather robust upper and lower boundaries on nuclear symmetry energies are extracted from the observational constraints up to about twice the saturation density $\rho_0$ of nuclear matter. More quantitatively, the symmetry energy at $2\rho_0$ is constrained to $E_{\rm{sym}}(2\rho_0)=46.9\pm10.1$ MeV excluding many existing theoretical predictions scattered between $E_{\rm{sym}}(2\rho_0)=15$ and 100 MeV. Moreover, by studying variations of the causality surface where the speed of sound equals that of light at central densities of the most massive neutron stars within the restricted EOS parameter space, the absolutely maximum mass of neutron stars is found to be 2.40 M$_{\odot}$ approximately independent of the EOSs used. This limiting mass is consistent with findings of several recent analyses and numerical general relativity simulations about the maximum mass of the possible super-massive remanent produced in the immediate aftermath of GW170817.
deformability
}
\authorrunning{Nai-Bo Zhang and Bao-An Li}
\titlerunning{High-Density Symmetry Energy from Neutron Stars and Gravitational Waves}
\maketitle
\section{Introduction}\label{intro}
To understand the nature and constrain the Equation of State (EOS) of dense neutron-rich matter in neutron stars (NSs) are a major goal shared by both nuclear physics and
astrophysics \cite{Dan02,Lat16,Watts16,Oertel17,Oz16a,Li17,David18}.
The nucleon specific energy $E(\rho ,\delta )$ in nucleonic matter of density $\rho$ and isospin asymmetry $\delta\equiv (\rho_n-\rho_p)/\rho$ (where $\rho_n$ and $\rho_p$ are densities of neutrons and protons, respectively) is a basic input for calculating the EOS of NS matter. Based on very extensive studies within essentially all existing nuclear many-body theories, the $E(\rho ,\delta )$ can be well approximated with the so-called empirical parabolic
law, see, e.g., ref. \cite{Bom91}
\begin{equation}\label{EOS0}
E(\rho ,\delta )=E_0(\rho)+E_{\rm{sym}}(\rho )\cdot\delta ^{2} +\mathcal{O}(\delta^4).
\end{equation}
The $E_0(\rho)$ is the energy in symmetric nuclear matter (SNM) of equal numbers of neutrons and protons while the $E_{\rm{sym}}(\rho )$ is generally called the symmetry energy of isospin asymmetric nuclear matter (ANM).

\begin{figure*}
\begin{center}
\resizebox{1\textwidth}{!}{
 \includegraphics{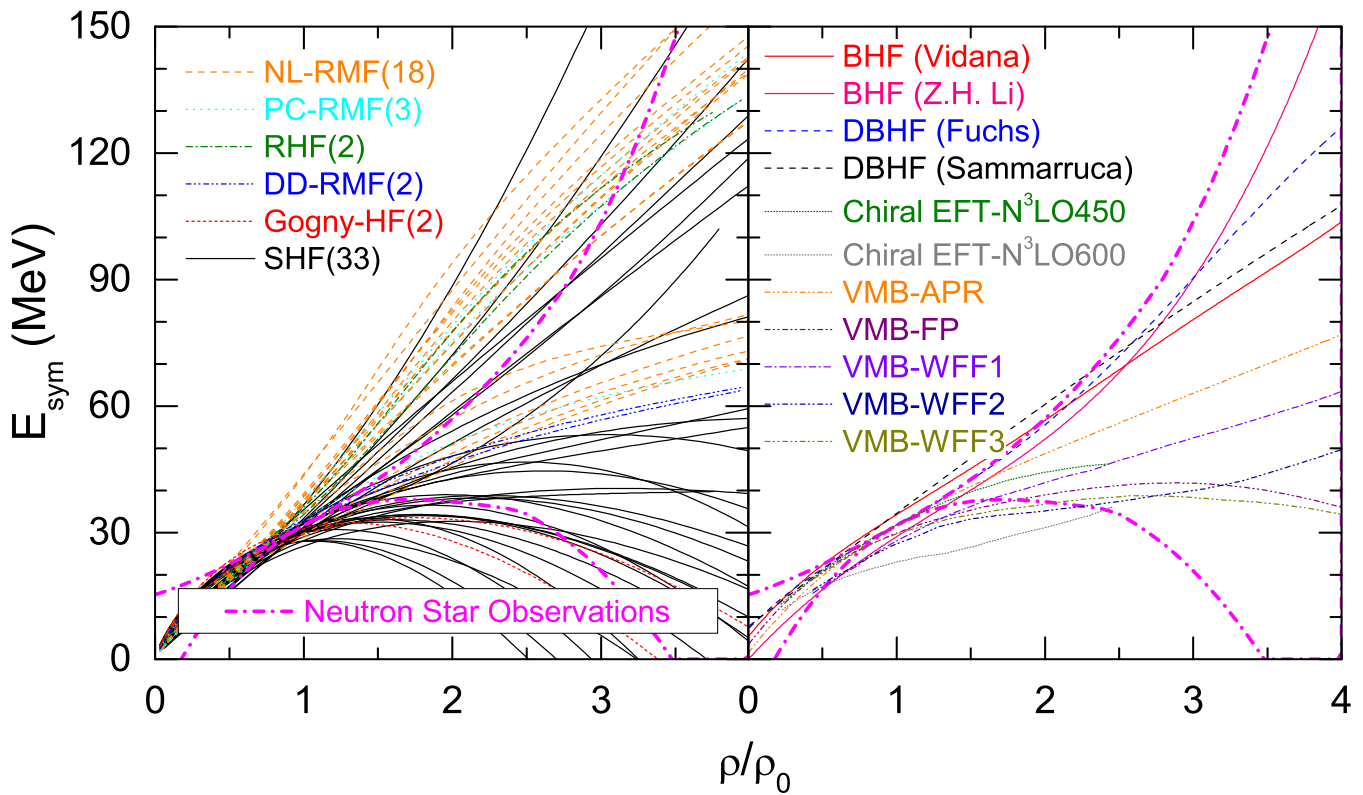}
}
\caption{(Color online) Examples of the density dependences of nuclear symmetry energy predicted by various kinds of nuclear many-body theories using different interactions, energy density functionals and/or techniques
(made by amending a compilation in ref. \cite{LWC17}) in comparison with the constraining boundaries (magenta dot-dashed lines) extracted in this work from studying properties of neutron stars.}
\label{Esym-survey}
\end{center}
\end{figure*}

While much progress has been made over the last few decades in obtaining both a theoretical understanding and observational/experimental constraints of the $E_0(\rho)$ as well as $E_{\rm{sym}}(\rho )$ around but mostly below the saturation density $\rho_0$ of nuclear matter \cite{ireview98,Bar05,Ste05,Li08,Tra12,Tsang12,Tesym,Hor14,Bal16}, even the trend of $E_{\rm{sym}}(\rho )$ at supra-saturation densities is presently controversial.
Since it is such an important quantity for investigating many issues in both astrophysics and nuclear physics, essentially all available nuclear many-body theories and interactions have been used to calculate the $E_{\rm{sym}}(\rho )$. While many calculations based on well-known theories predict an increasing $E_{\rm{sym}}(\rho )$ with density, a large number of equally well-respected theories predict that the $E_{\rm{sym}}(\rho )$ first increases with density passing $\rho_0$, then stays approximately a constant (or decreases above certain densities) depending on the isospin-dependence of the short-range tensor or three-body nuclear force used in the theories. Therefore, the $E_{\rm{sym}}(\rho )$ at high densities has been broadly recognized as the most uncertain part of the EOS of dense neutron-rich nucleonic matter \cite{Tesym}. For example, over 520 nuclear energy density functionals available by 2014 have been used to predict the $E_{\rm{sym}}(\rho )$, see, e.g., refs. \cite{Roc11,Dutra1,Dutra2}.  Shown in the left window of
Fig. \ref{Esym-survey} are 60 selected representatives from 6 classes of phenomenological models and/or energy density functional theories including the Relativistic Mean Field (RMF) using 3 different kinds of coupling schemes, Relativistic Hartree-Fock (RHF), Gogny Hartree-Fock (HF) and Skyrme Hartree-Fock (for a detailed list of the interactions/models used, see the compilation by Lie-Wen Chen in ref. \cite{LWC17}).
The large spread in the predicted symmetry energies especially at high densities clearly calls for experimental/observational constraints. To our best knowledge, all microscopic and/or {\it ab initio} theories have also been used to predict the $E_{\rm{sym}}(\rho )$. Shown in the right window of Fig. \ref{Esym-survey} are 11 examples \cite{Vid09,LiZ08,Kla06,Sam10,APR98,Fri81,WFF12,Fra14}. They are from the Brueckner Hartree Fock (BHF), Dirac-Brueckner Hartree Fock (DBHF), Chiral Effective Field Theory (Chiral EFT) and the Variational Many Body (VMB) theory using different interactions and/or high-momentum cut-offs. Their predictions also spread broadly at supra-saturation densities. In fact, by design, some of these microscopic theories are valid only at low-energies/densities. When they are extrapolated to high densities, their predictions may not converge and often depend on the high-momentum cut-off used in the theories. A useful measure of the predicted spread of high-density symmetry energies is the value of symmetry energy at twice the saturation density $E_{\rm{sym}}(2\rho_0)$. Information about the EOS and symmetry energy around this density is most relevant for determining the radii of NSs \cite{Lattimer01} and heavy-ion reactions with radioactive beams of about 400 MeV/nucleon \cite{Li08}. The examples shown in Fig. \ref{Esym-survey} have $E_{\rm{sym}}(2\rho_0)$ values scatter between approximately $15$ to $100$ MeV \cite{LWC15}. The magenta dot-dashed lines are the boundaries of $E_{\rm{sym}}(\rho )$ we extracted in this work from studying properties of NSs as we shall explain in detail in the following. Clearly, the extracted constraint on symmetry energy can already exclude many of the predictions while it is still quite loose at densities above $2\rho_0$.

The proton fraction $x_p(\rho)$ in NSs at $\beta$-equilibrium is uniquely determined by the $E_{\rm{sym}}(\rho )$. Consequently, the composition, critical nucleon density $\rho_c$ (where $x_p(\rho_c)=1/9$) above which the fast cooling by neutrino emissions through the direct URCA process can occur, and the crust-core transition density in NSs all depend sensitively on the $E_{\rm{sym}}(\rho )$. It is well known that the radii of NSs are most sensitive to the pressure around $1-2\rho_0$ where the symmetry energy makes a significant contribution to the pressure \cite{Lattimer01}.  Moreover, the frequencies and damping times of various oscillations, quadrupole deformations of isolated NSs and the tidal deformability in NS mergers also depend on the $E_{\rm{sym}}(\rho )$ \cite{Newton2}. Furthermore, there is a degeneracy between the EOS of super-dense neutron-rich matter and the strong-field gravity in understanding both properties of super-massive NSs and the minimum mass to form black holes. Thus, further testing Einstein's General Relativity (GR) against alternative theories of super-strong gravity also requires reliable knowledge about the EOS of dense neutron-rich matter, especially its $E_{\rm{sym}}(\rho )$ term at supra-saturation densities \cite{DeHua1,XTHE}.

The fundamental origin of uncertainties of the high-density $E_{\rm{sym}}(\rho )$ is the largely unknown and relatively weak isospin-dependence (i.e., the difference between neutron-proton interactions in the isosinglet and
isotriplet channels, while neutron-proton, neutron-neutron and proton-proton interactions are the same in the isotriplet channel due to charge independence) of nuclear forces especially at short distances in the dense neutron-rich medium \cite{PPNP}. Determining the high-density $E_{\rm{sym}}(\rho )$ using nuclear reactions induced by high-energy rare isotope beams has been identified as a major science thrust in both the 2015 American and 2017 European nuclear physics long range plans for the next decade \cite{LRP1,LRP2}. Unlike the small isospin effects in laboratory experiments with finite nuclei of normally small neutron/proton ratios, NSs are the natural testing ground of fundamental interactions at extremely high densities and/or isospin asymmetries in cold matter.

In this work, using an explicitly isospin-dependent parametric EOS in terms of the $E(\rho ,\delta )$ of neutron-rich nucleonic matter with its two components $E_0(\rho)$ and $E_{\rm{sym}}(\rho )$ both parameterized as functions of density up to $(\rho/\rho_0)^3$ terms respecting available constraints around $\rho_0$,  we invert the TOV equation to investigate how the observational constraints on the radii, maximum mass and tidal deformability of NSs as well as the causality requirement all together may constrain the high-density $E_{\rm{sym}}(\rho)$.  Rather strong upper and lower bounds on the $E_{\rm{sym}}(\rho)$ at supra-saturation densities are obtained. In particular, the symmetry energy at $2\rho_0$ is constrained to E$_{\rm{sym}}(2\rho_0)=46.9\pm10.1$ MeV excluding many of the existing predictions as shown in Fig. \ref{Esym-survey}.
Simultaneously, the total pressure $P(\epsilon)$ as a function of energy density $\epsilon$ is constrained into a band consistent with that extracted recently by the LIGO+Virgo Collaborations from their improved analyses of the NS tidal deformability in GW170817 \cite{LIGO2018}. Moreover, by studying variations of causality surface within the restricted high-density EOS parameter space, the absolutely maximum mass of neutron stars is found to be 2.40 M$_{\odot}$ almost independent of the EOSs used. The limiting mass for NSs is consistent with recent findings from studying the electromagnetic signals and numerical GR simulations of GW170817 \cite{Bau17,Margalit17,Shibata17,Rezzolla18,Ruiz18,Radice18}.

The paper is organized as follows. In the next section, we first outline the approach of numerically inverting the TOV equation using the parametric EOS first used in our previous work \cite{NBZ2018a}. We then study in sect. \ref{cau} causality constraints on the maximum mass of NSs and the EOS parameter space. In sect. \ref{obs}, observational limits on parameters characterizing the high-density EOS are examined. These limits are then transformed
into constraints on the pressure $P(\epsilon)$ at $\beta$ equilibrium in NSs and the symmetry energy $E_{\rm{sym}}(\rho )$ in sect. \ref{com1} and sect. \ref{com2}, respectively.  A summary is given at the end.
\begin{figure*}
\begin{center}
\resizebox{0.99\textwidth}{!}{
  \includegraphics{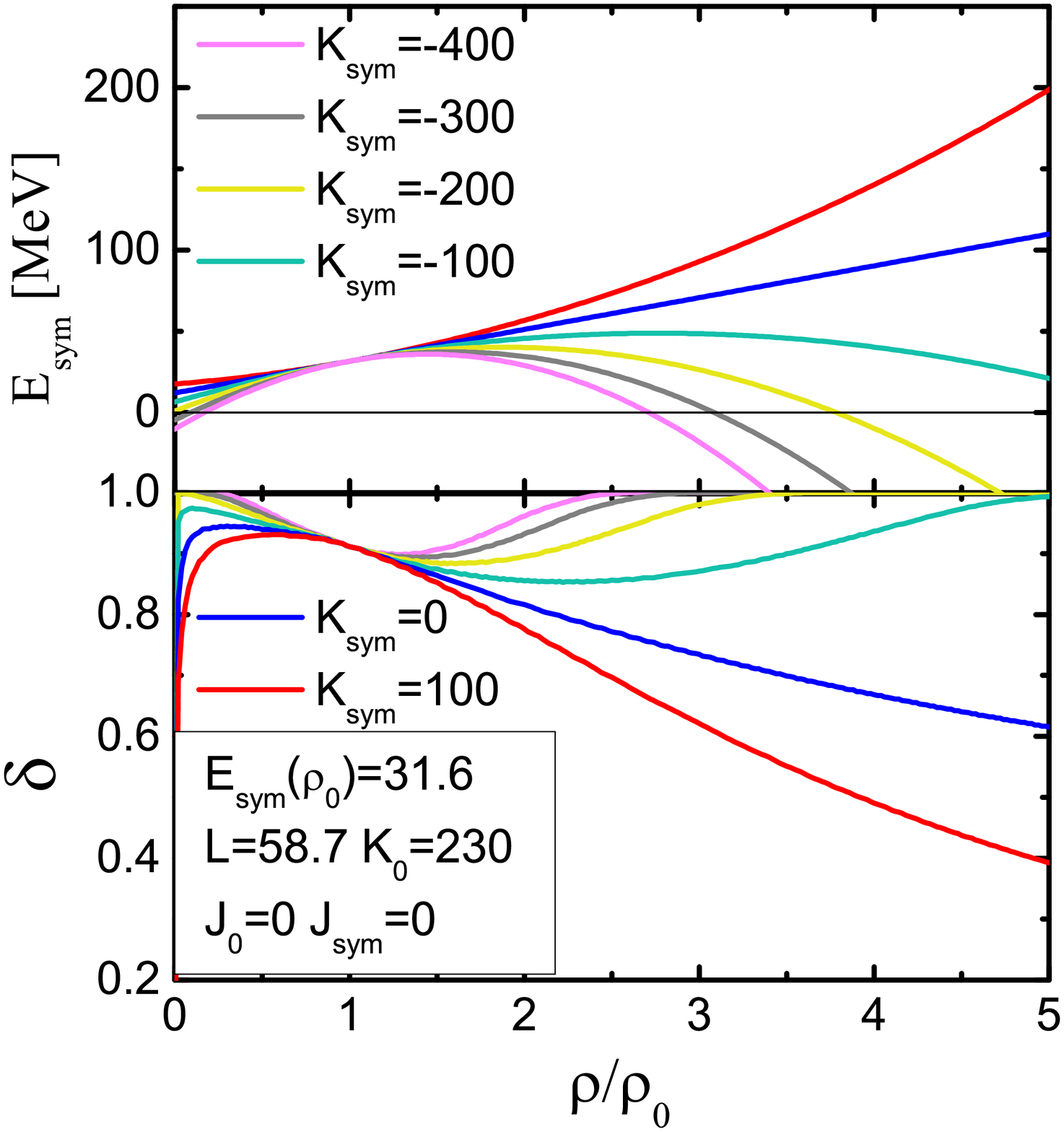}
  \includegraphics{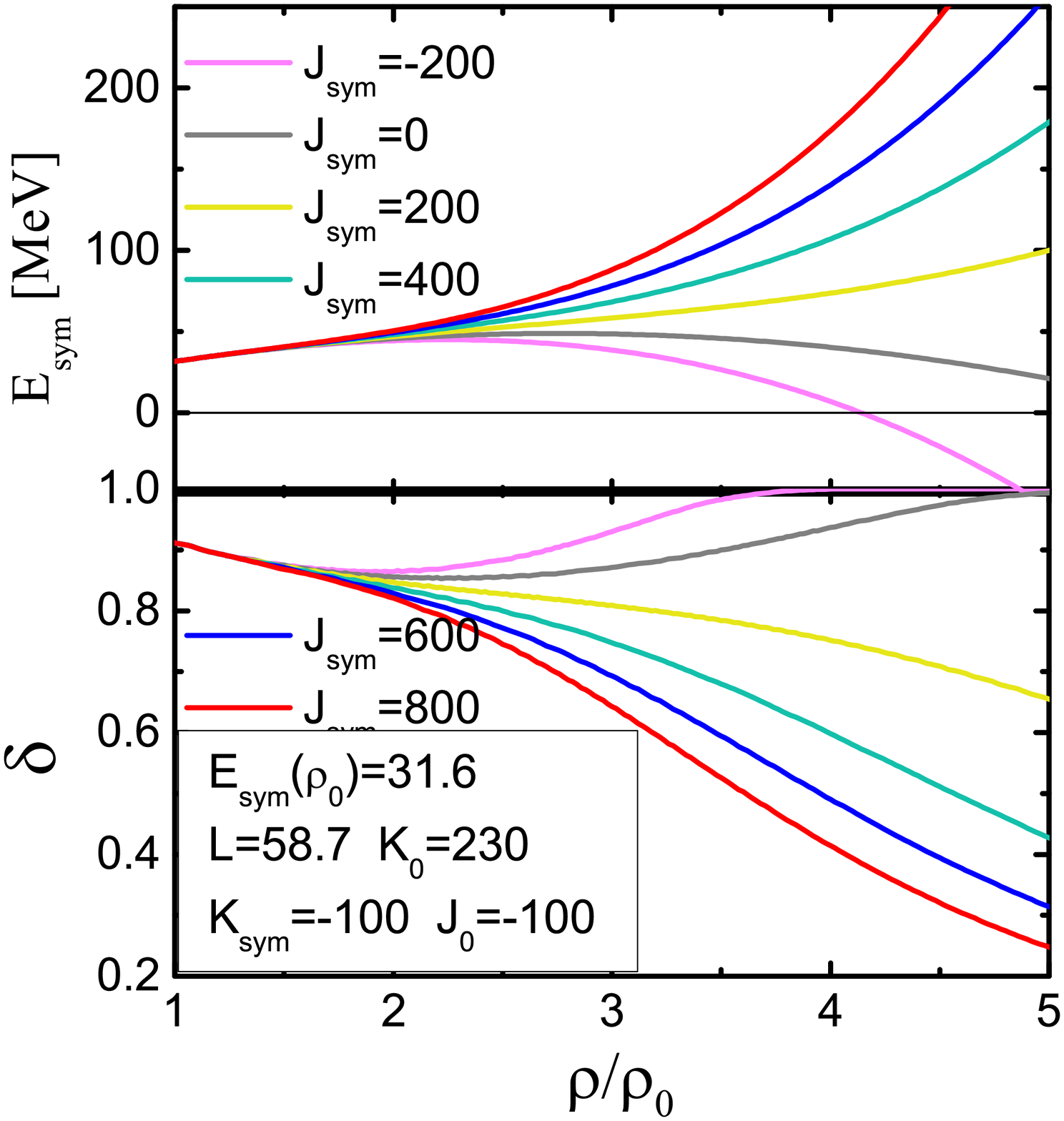}
  }
  \caption{(color online) The symmetry energy $E_{\rm{sym}}(\rho)$ and isospin asymmetry $\delta(\rho)$ in neutron star matter at $\beta$-equilibrium as a function of the reduced density $\rho/\rho_0$ for $K_{\rm{sym}}=-400$, -300, -200, -100, 0, and 100 MeV (left), and $J_{\rm{sym}}=-200$, 0, 200, 400, 600, and 800 MeV (right), respectively, while the fixed parameters at the saturation density are specified. All parameters are in unit of MeV. Taken from ref. \cite{NBZ2018a}}\label{Ksymeffect}
  \end{center}
\end{figure*}

\section{The theoretical framework}\label{tov}
It is well known that the mass (M)-radius (R) relationship of neutron stars can be obtained by solving the TOV equation \cite{Oppenheimer39}
\begin{equation}\label{TOVp}
\frac{dP}{dr}=-\frac{G(m(r)+4\pi r^3P/c^2)(\epsilon+P/c^2)}{r(r-2Gm(r)/c^2)},
\end{equation}
\begin{equation}\label{TOVm}
\frac{dm(r)}{dr}=4\pi\epsilon r^2
\end{equation}
for a given input EOS $P(\epsilon)$ where $G$ is the gravitation constant, $c$ is the light speed and $m(r)$ is the gravitational mass enclosed within a radius $r$.

The quadrupole tidal deformations of NSs due to their mutual gravity during the inspiring phase of their mergers are known to be sensitive to the EOS. 
They affect the phase angles of GW waveforms. The tidal deformability
inferred from GW signals can thus be used to constrain the EOS of dense neutron-rich matter.
It is customary to use the {\it dimensionless} tidal deformability
\begin{equation}
\Lambda = \frac{2}{3}k_2\cdot (Rc^2/GM)^5
\end{equation}
where $k_2$ is the Love number determined by the EOS through a differential equation coupled to the TOV equation \cite{Hinderer08,Hinderer10}. It is particularly interesting to note that the tidal deformability
was found sensitive to the high-density behavior of nuclear symmetry energy but not much to the saturation properties of nuclear matter \cite{Farr1,Farr2}.

Since the main purpose of our work is to constrain the EOS of neutron-rich matter and extract the underlying nuclear symmetry energy $E_{\rm{sym}}(\rho)$ using NS observations,
we shall invert numerically the TOV equation for a given single or set of NS observables using a parametric EOS with explicit isospin dependence. Detailed discussions of this approach are presented
in our previous work \cite{NBZ2018a}.  For completeness and ease of our following discussions, we summarize here the main features and justifications of our approach.

\subsection{Why we need to go beyond the traditional way of parameterizing directly the high-density pressure in neutron stars using piecewise polytropes}
The usual representations of the high-density EOS of NS matter widely used in the literature are based on parameterizing the adiabatic index $\Gamma(P)$
defined by
$
\Gamma(P) = \frac{\epsilon+P}{P}\frac{dP}{d\epsilon}
$
with pieceweise analytical functions in each of n density/pressure domains \cite{Lee00}, e.g., the piecewise polytropes from using constant $\Gamma$ values in the n domains, or from constructing Log$(\Gamma(P))$ as a polynomial of Log$(P)$, for a recent review, see, e.g., ref. \cite{Lee2018}. While these types of parametric EOSs are indeed sufficient for solving the TOV equation and have been found very useful for many purposes, they carry no non-degenerate information about the internal composition of dense matter and are unable to reveal clearly the  underlying nuclear symmetry energy. To extract directly information about the high-density nuclear symmetry energy from inverting the TOV equation using observational constraints, one thus has to parameterize the EOS at a more basic level using the isospin degree of freedom explicitly according to Eq. (\ref{EOS0}).

It is necessary to note here that there are accurate predictions for the EOS of pure neutron matter (PNM) up to about $1.3\rho_0$ based on microscopic and/or ab initio nuclear theories, see, e.g., ref. \cite{Ste15} for a recent review. These predictions have been widely used to calibrate the EOS of neutron-rich matter mostly at very low densities, see, e.g., \cite{Newton12,Fatt12}. Indeed, the NS EOSs constructed from extrapolating the PNM EOS to high densities have been found very useful in understanding some NS observables, see, e.g., refs. \cite{Heb10,Ste12,Tews17}. One may show that the total energy of NS matter around $1-2\rho_0$ can be well approximated by that of PNM for some purposes, i.e, approximating the nucleon specific energy in PNM as the nuclear symmetry energy by neglecting the EOS of SNM, which is equivalent to approximating the total pressure from observations as completely due to the symmetry energy and considering NSs as made of only PNM. Thus, within this approximation one can also obtain some useful information about the high-density symmetry energy from the polytropes extracted from analyzing astrophysical observations. We note, however, numerical calculations have shown that the relative contributions of symmetry energy and SNM EOS to the total pressure of NS matter at $\beta$ equilibrium depend strongly on the density dependence of the symmetry energy in the $1-2\rho_0$ density range, see, e.g., results shown in Fig. 146 of ref. \cite{Li08}. Moreover, detailed information about the proton fraction in this density range, albeit possibly small, is critical for understanding the cooling mechanism of protoneutron stars. Thus, a more accurate and consistent method of extracting explicitly the nuclear symmetry energy at supra-saturation densities directly from astrophysical observations is absolutely necessary. While our approach used in this work is certainly not perfect, it helps overcome the difficulties of knowing explicitly the internal composition of NS matter without considering the isospin degree of freedom in approaches using the polytropes to parameterize directly the EOS at high densities.

\subsection{An explicitly isospin-dependent parametric EOS for high-density neutron star matter}
The energy density $\epsilon(\rho, \delta)$ in neutron star matter is
\begin{equation}\label{energydensity}
  \epsilon(\rho, \delta)=\epsilon_N(\rho, \delta)+\epsilon_l(\rho, \delta),
\end{equation}
where $\epsilon_N(\rho, \delta)$ and $\epsilon_l(\rho, \delta)$ are the energy density of nucleons and leptons, respectively. The $\epsilon_N(\rho, \delta)$ is determined by the
nucleon specific energy $E(\rho,\delta)$ and average mass $M_N$ via
\begin{eqnarray}\label{Ebpa}
  \epsilon_N(\rho,\delta)=\rho E(\rho,\delta)+\rho M_N.
\end{eqnarray}
The pressure in neutron star matter is then calculated from
\begin{equation}\label{pressure}
  P(\rho, \delta)=\rho^2\frac{d\epsilon(\rho,\delta)/\rho}{d\rho}.
\end{equation}
The pressure $P(\rho,\delta)$ and energy density $\epsilon(\rho,\delta)$ become functions of density only once the isospin asymmetry profile $\delta(\rho)$ is determined from
 the $\beta$-equilibrium condition
$
  \mu_n-\mu_p=\mu_e=\mu_\mu\approx4\delta E_{\rm{sym}}(\rho)
$
and the charge neutrality condition
$  \rho_p=\rho_e+\rho_\mu.
$
The chemical potential for a particle $i$ can be calculated from
$
  \mu_i=\partial\epsilon(\rho,\delta)/\partial\rho_i.
$

Once the electron chemical potential becomes larger than the muon rest mass, muons start to appear. The critical density for muon formation is calculated consistently.
The lepton energy density $\epsilon_l(\rho, \delta)$ is calculated from the noninteracting Fermi gas model as normally done in the literature \cite{Oppenheimer39}.
As discussed in detail in our previous work \cite{NBZ2018a}, we prepare self-consistently the EOS of $npe\mu$ matter in NSs. Nevertheless, it is instructive to write down here the pressure of $npe$ matter before muons appear 
\begin{equation}\label{pre-npe}
  P(\rho, \delta)=\rho^2[E^{'}_0(\rho)+E^{'}_{\rm{sym}}(\rho)\delta^2]+\frac{1}{2}\delta (1-\delta)\rho E_{\rm{sym}}(\rho). 
\end{equation}
It shows analytically the roles of both the SNM EOS and nuclear symmetry energy in determining the pressure of $npe$ matter at $\beta$ equilibrium. The isospin symmetric and asymmetric parts of the above pressure 
using several typical density dependences of nuclear symmetry energy were compared in refs. \cite{Li08,Li-mex}.  We shall quote those results in our qualitative discussions regarding the relative effects of SNM EOS and symmetry energy on the pressure in NSs around twice the saturation density of nuclear matter in Sec. \ref{comparison}.

\begin{figure*}
\begin{center}
\resizebox{0.49\textwidth}{!}{
  \includegraphics{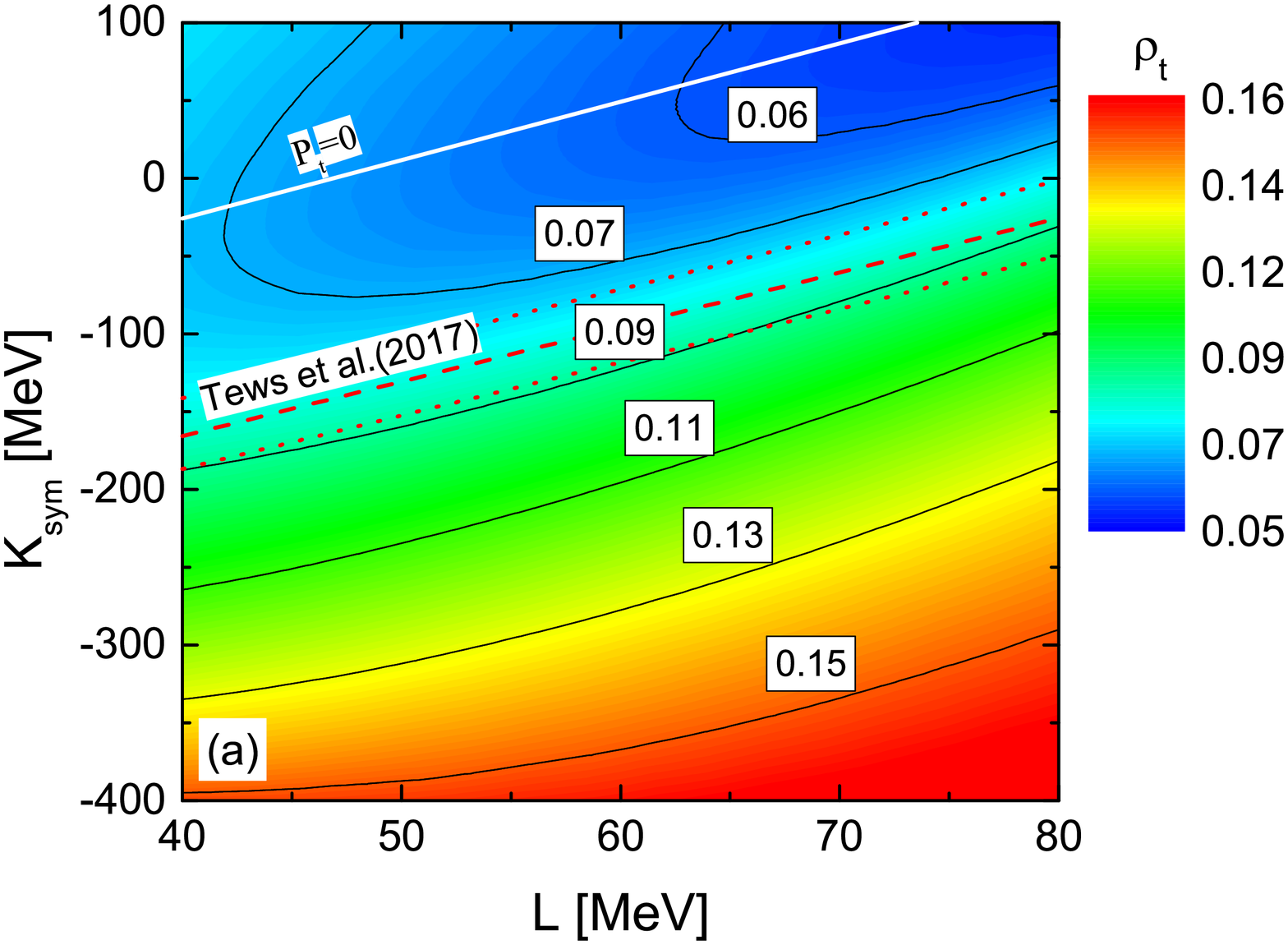}
  }
  \resizebox{0.49\textwidth}{!}{
  \includegraphics{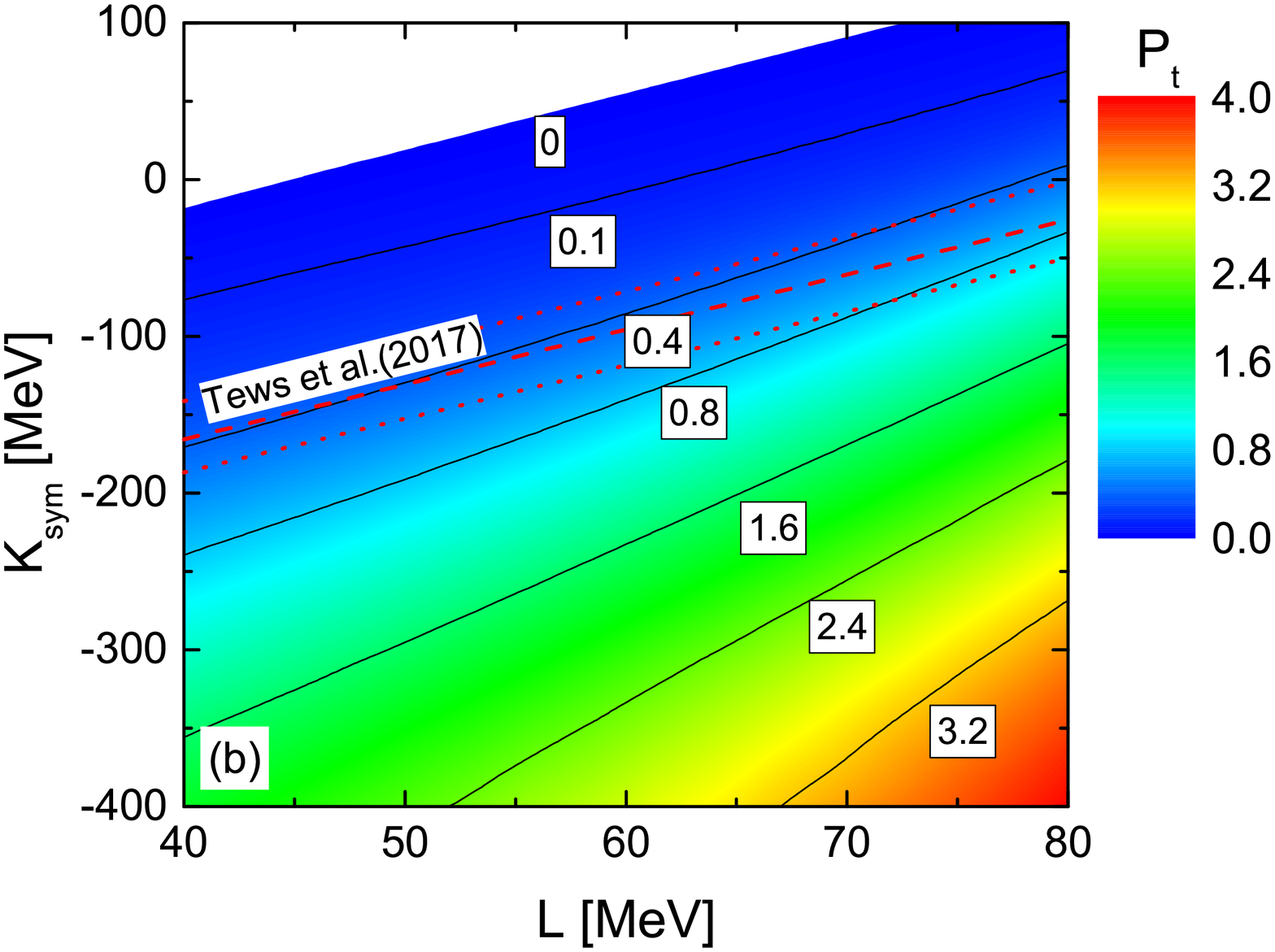}
}
\end{center}
  \caption{(color online)
 Contours of the crust-core transition density $\rho_t$ in fm$^{-3}$ (a) and the corresponding pressure $P_t$ in MeV fm$^{-3}$ (b) in the $L-K_{\rm{sym}}$ plane. Lines with fixed values of transition densities and pressures are labeled. The red dashed lines are the correlations between $K_{\rm{sym}}$ and $L$ from \cite{Tews17}. The white line in (a) is where the transition pressure vanishes. Modified from two figures in ref. \cite{NBZ2018a}.}\label{Trandp}

\label{obsers}       
\end{figure*}

To infer the high-density nuclear symmetry energy from NS observables, it is necessary to start from directly parameterizing the $E_0(\rho)$ and $E_{\rm{sym}}(\rho )$ in Eq. (\ref{EOS0}), separately.
Since the density profile of isospin asymmetry $\delta(\rho)$ in charge neutral NS matter at $\beta$ equilibrium is completely determined by the $E_{\rm{sym}}(\rho)$,  the corresponding $P(\epsilon)$ can then be calculated consistently once the $E_0(\rho)$ and $E_{\rm{sym}}(\rho )$ are known. We parameterize them as
\begin{eqnarray}\label{E0para}
E_{0}(\rho)&=&E_0(\rho_0)+\frac{K_0}{2}(\frac{\rho-\rho_0}{3\rho_0})^2+\frac{J_0}{6}(\frac{\rho-\rho_0}{3\rho_0})^3,\\
E_{\rm{sym}}(\rho)&=&E_{\rm{sym}}(\rho_0)+L(\frac{\rho-\rho_0}{3\rho_0})+\frac{K_{\rm{sym}}}{2}(\frac{\rho-\rho_0}{3\rho_0})^2\nonumber\\
&+&\frac{J_{\rm{sym}}}{6}(\frac{\rho-\rho_0}{3\rho_0})^3.\label{Esympara}
\end{eqnarray}
It is useful to emphasize that the above parameterizations naturally approach asymptotically their Taylor expansions when $\rho\rightarrow \rho_0$.  Thus, as discussed in great detail in ref. \cite{NBZ2018a}, the above two expressions have the dual meaning of being Taylor expansions near the saturation density but parameterizations at supra-saturation densities. While leaving the high-density parameters $J_0$, $K_{\rm{sym}}$ and
$J_{\rm{sym}}$ as free parameters to be determined by NS observables, we fix the $E_0(\rho_0),K_0, E_{\rm{sym}}(\rho_0)$ and L at their currently known most probable values at saturation density.
Unlike in some other approaches, such as in energy density functional analyses where the above expressions are used as Taylor expansions and the coefficients are normally correlated, 
we treat all the coefficients as independent parameters. In nowhere we try to predict the high-density parameters from the low-density ones. In short, especially when applied to high densities, the above two expressions are not expansions or extrapolations but simply parameterizations. 

We are here trying to solve the inverse-structure problem of NSs, namely to infer from the observations parameters of the EOS. The parameterizations are used as tools. Indeed, there is 
a genuine question: does the results of our study depend on the tools we use? For example, what happens if one uses more terms in parameterizing the EOS of SNM and the symmetry energy?  Firstly, given the remaining uncertainties of the currently used high-density parameters and the limited data available, there is currently no observational/experimental information we can use to constrain more parameters than we already used in the above two equations. Secondly, the above parameterizations are already flexible enough to mimic diverse behaviors of the EOS of SNM and the symmetry energy predicted by various theories in the literature. 
In solving the NS inverse-structure problem, we expect that our results on the extracted symmetry energy at high densities to be independent of the parameterizations we used while we have not proved this yet. 
Thus, similar to some existing studies about effects of using different segments and forms of polytropes on extracting pressures at high densities in NSs, it would be interesting to study how using different parameterizations of the EOS of SNM and nuclear symmetry energy in the same high-density region may affect what we 
extract from the same observational data. Such a study is being planned.

Extensive studies over the last four decades have determined the most probable incompressibility of symmetric nuclear matter as $K_0=230 \pm 20$ MeV \cite{Shlomo06,Piekarewicz10},  while surveys of 53 analyses done over the last two decades have found the most probable magnitude $E_{\rm sym}(\rho_0)=31.7\pm 3.2$ MeV and slope $L=58.7\pm 28.1 $ MeV of symmetry energy at $\rho_0$ \cite{Oertel17,Li13}. There are also theoretical and experimental efforts to determine the high-density EOS parameters. However, they are still very uncertain \cite{Tews17,Zhang17}.
Nevertheless, the currently known ranges of $-400 \leq K_{\rm{sym}} \leq 100$ MeV, $-200 \leq J_{\rm{sym}}\leq 800$ MeV, and $-800 \leq J_{0}\leq 400$ MeV provide at least a starting point in our efforts to narrow them down using NS observables. Once these parameters are constrained with NS observables, the $E_0(\rho)$, $E_{\rm{sym}}(\rho )$ and subsequently the $P(\epsilon)$ can all be reconstructed.
This prior information albeit inaccurate about the ranges of the high-density EOS parameters is useful compared to directly parameterizing the $P(\epsilon)$ for which there is even less accurate prior information available.

As an illustration of the high-density behaviors of the $E_{\rm{sym}}(\rho)$ to be explored by varying the $K_{\rm{sym}}$ and $J_{\rm{sym}}$ parameters, shown in Fig. \ref{Ksymeffect} are the symmetry energy $E_{\rm{sym}}(\rho)$ and the resulting isospin asymmetry profile $\delta(\rho)$ as functions of the reduced density $\rho/\rho_0$ by varying only one parameter each time while fixing all others. In the left window, we used $K_{\rm{sym}}=-400$, -300, -200, -100, 0, and 100 MeV, and in the right window $J_{\rm{sym}}=-200$, 0, 200, 400, 600, and 800 MeV, respectively. As their names indicate, the slope $L$, curvature $K_{\rm{sym}}$ and skewness  $J_{\rm{sym}}$ of symmetry energy play different roles and in order become increasingly more important at higher densities. Clearly, very diverse density dependences of the $E_{\rm{sym}}(\rho)$ spanning the whole range of model predictions shown in Fig. \ref{Esym-survey} are sampled. The resulting $\delta(\rho)$ at $\beta$ equilibrium shown in the lower part of Fig. \ref{Ksymeffect} varies from values for very neutron-poor matter with
stiff symmetry energies to pure neutron matter when the $E_{\rm{sym}}(\rho)$ becomes zero or even negative.

\subsection{The crust-core transition and limits of the EOS parameters to ensure the thermodynamical stability of NSs}
Lindblom recently reminded us several basic guiding principles that all parameterized EOSs should follow \cite{Lee2018}. These include (1) any EOS should be representable by an appropriate parametric representation to any level of desired accuracy, (2) each representation should satisfy the basic laws of physics including the thermodynamic stability and causality, 3) the representation should be efficient. While varying the three high-density EOS parameters within our approach outlined above, necessary physics measures are taken to ensure that our EOS representation meets these requirements. In particular, the entire EOS from the surface to the core is ensured to be thermodynamically stable. 

To connect self-consistently the core EOS described above with the NV EOS \cite{Negele73} for the inner crust and the BPS EOS  \cite{Baym71} for the outer crust, the core-crust transition density 
was found by examining the incompressibility of NS matter for any given set of EOS parameters \cite{NBZ2018a}. In addition, to ensure the thermodynamical stability of NSs, we require the transition pressure to
stay positive. This condition limits the low-density behavior of nuclear symmetry energy and is useful to restrict the EOS parameter space. As an example, shown in the left and right window of Fig. \ref{Trandp} are contours of
the transition density and pressure, respectively, in the $K_{\rm{sym}}$-$L$ plane with other relevant parameters set at their most probable values at saturation density discussed earlier. The transition density and pressure 
depend more strongly on $K_{\rm{sym}}$ than $L$ in their uncertainty ranges considered. As a parameter affecting only the behavior of the symmetry energy at densities above about $2.5\rho_0$, the $J_{\rm{sym}}$ has no effect on the crust-core transition.  For transition densities higher than about $\rho_t=0.07$ fm$^{-3}$, the required $K_{\rm{sym}}$ for a given constant $\rho_t$ increases monotonically with $L$. For a fixed $P_t$,  the required $K_{\rm{sym}}$ always increases linearly with $L$ before reaching the stability boundary $P_t=0$ along the line $K_{\rm{sym}}=3.64 L-163.96~(\rm{MeV})$. Depending on the value of $L$, some regions of the $K_{\rm{sym}}$ parameter space (the white area above the line labeled with $P_t=0$ in the right window) is excluded. The corresponding very low transition densities are also excluded as indicated by the white line in the left window. 

While we use the $K_{\rm{sym}}$ and $L$ as independent parameters in our approach, it is interesting to see how the empirical correlation between the $K_{\rm{sym}}$ and $L$ from analyzing systematically predictions using various many-body theories and interactions limits the crust-core transition properties. For example, Tews et al \cite{Tews17} found the relation 
$K_{\rm{sym}}=3.501L-305.67\pm24.26~\rm{(MeV)}$. This correlation together with its 68.3\% lower and upper boundaries are shown with the red dashed lines, restricting the $\rho_t$ and $P_t$ to around $0.08$ fm$^{-3}$ and $0.4$ MeV fm$^{-3}$, respectively. These values are consistent with the crust-core transition properties often used in the literature. For the purposes of this work to solve the NS inverse-structure problem, however, we do not use any predicted correlations among the EOS parameters. Instead we vary the parameters independently in constructing self-consistently the entire EOS from the surface to core of NSs and study how the observational properties may limit the EOS parameter space. In the present study focusing on extracting the high-density symmetry energy from observations of neutrons and gravitational waves, as mentioned earlier, the value of $L$ is fixed at its most probable value of 58.7 MeV, the crust-core transition density and pressure therefore only change with $K_{\rm{sym}}$ in the ranges shown in Fig. \ref{Trandp}.
 
As we have shown earlier, the combined variations of the three high-density parameters are flexible enough for us to mimic essentially all kinds of high-density EOSs predicted by various many-body theories. Of course, we limited ourselves to the EOSs for $npe\mu$ matter in the minimum NS model without phase transitions. As to the causality, we actually use it to limit the allowed parameter space, thus the high-density EOS. As we shall discuss next, reaching the causal limit at central densities of the most massive neutron stars is used to determine the absolutely maximum mass of NSs.

\section{Causality limits on the absolutely maximum mass and central density of neutron stars as well as the high-density symmetry energy parameters}\label{cau}
\begin{figure}
\begin{center}
\resizebox{0.45\textwidth}{!}{
  \includegraphics{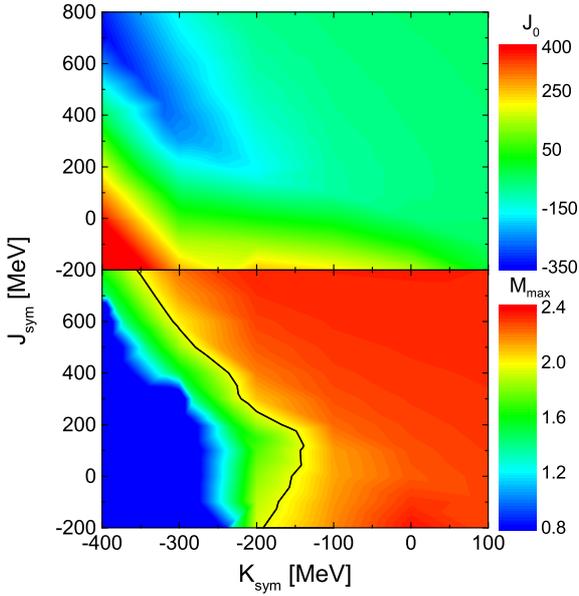}
}
\end{center}
  \caption{(color online) Contours of the $J_0$ parameter and the corresponding mass M$_{\rm{max}}$ of the most massive neutron stars on the causality surface in the plane of $J_{\rm{sym}}$ versus $K_{\rm{sym}}$. The solid line in the lower window is the boundary where M$_{\rm{max}}=2.01$ M$_{\odot}$. }
\label{cau1}       
\end{figure}
The speed of sound is defined by $v^2_s=dP/d\epsilon$. The thermodynamical stability and causality condition require that $0\leq v^2_s\leq c^2$. The latter limits naturally the EOS parameter space and determines the absolutely maximum mass of NSs. To use the causal limit for our purposes, we search for combinations of the three high-density EOS parameters leading to $v^2_s=c^2$ right at the central density of the most massive NS supported by a given EOS from solving the TOV equation. We refer the surface where the above conditions are met in the 3D ($J_0-J_{\rm{sym}}-K_{\rm{sym}}$) EOS parameter space as the causality surface in our following discussions.

\subsection{Causality contour in the parameter plane of high-density nuclear symmetry energy}
First, let us examine the contour of the $J_0$ parameter in the plane of $J_{\rm{sym}}$ versus $K_{\rm{sym}}$ where the causal limit has been reached at the central density of the most massive NS configuration
supported by the individual EOS specified by the three high-density EOS parameters in the upper window of Fig. \ref{cau1}. The contour of the corresponding maximum mass is shown in the lower window. We begin from the left side. The lower left corner in the $J_{\rm{sym}}-K_{\rm{sym}}$ plane is where the nuclear symmetry energy is super-soft with both $J_{\rm{sym}}$ and $K_{\rm{sym}}$ being negative. The maximum mass of NSs in this area is less than 1.4 M$_{\odot}$. To support NS sequences with a maximum mass as low as M$_{\rm{max}}=0.8$ M$_{\odot}$ would require a $J_0$ as high as 400 MeV to make the total pressure strong enough. It clearly demonstrates the complementary contributions of the symmetry energy and the SNM EOS to the total pressure necessary to support stable NSs. Overall, in the left region where the maximum masses are low, the variation of the $J_0$ is more obvious, meaning that maximum masses of these lighter NSs are more sensitive to the variation of the EOS. On the contrary, in the big red region on the right in the lower window the maximum mass is almost a constant of about 2.4 M$_{\odot}$. Interestingly, the corresponding region in the upper window covers values of $J_0$ ranging from -350 to 400 MeV. Combing the information from both windows, it is seen that there is a big parameter space where the M$_{\rm{max}}=$2.4 M$_{\odot}$, namely this limiting value of the maximum mass is almost independent of the EOSs used. We refer it as the absolutely maximum mass (i.e., the maximum of the maximum masses using any EOS) of NSs. It will be further studied next by examining several other features of the causality surface.

While theoretically NSs may have masses up to a currently unknown maximum mass above which NSs will collapse into black holes, the two most massive NSs observed reliably so far are the
pulsar J1614-2230 with M=$1.97\pm0.04$ M$_\odot$ \cite{Demorest10} and the pulsar J0348+0432 with M=$2.01\pm0.04$ M$_\odot$ \cite{Antoniadis13}. Since all reasonable EOSs are required to
be stiff enough to support at least the most massive NSs observed so far, we refer the observed maximum mass of M$_{\rm{max}}=2.01$ M$_{\odot}$ as the minimum maximum mass of NSs in our following discussions.
This mass is shown as the solid line in the lower window of Fig. \ref{cau1}. The fine structure (wiggles) along this line is due to the bin sizes of the EOS parameters we used. The errors introduced by this finite bin size effect are
much smaller than those due to the uncertainties of the EOS parameters themselves. The region to its left is excluded by this observational constraint.
Observations of more massive NSs certainly will further limit the EOS parameter space. It is thus interesting to note that the mass of the neutron star PSRJ2215+5135 was very recently reported to be $2.27^{+0.17}_{-0.15} $ M$_\odot$ \cite{newM}. If confirmed, this will raise the minimum maximum mass of NSs, subsequently restrict further the allowed EOS parameter space.

\begin{figure}
\begin{center}
\resizebox{0.7\textwidth}{!}{
  \includegraphics{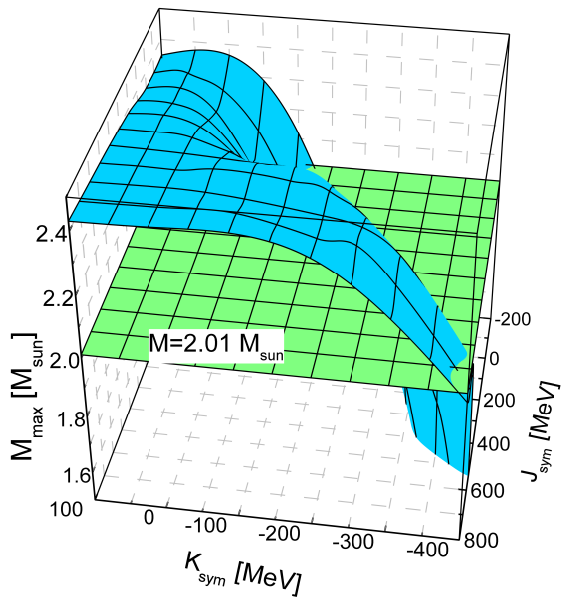}
  }
 \end{center}
  \caption{(color online) The mass of the most massive neutron stars on the causality surface as functions of
  $J_{\rm{sym}}$ and $K_{\rm{sym}}$, respectively. }
\label{mass1}       
\end{figure}

The same information about the causality surface and the associated physics may be more clearly visualized by presenting them in another way.
Shown in Fig. \ref{mass1} is the maximum mass of NSs on the causality surface as a function of $J_{\rm{sym}}$ and $K_{\rm{sym}}$ in 3D. For comparisons, a plane with the current minimum maximum mass M$_{\rm{max}}=2.01$ M$_{\odot}$ from the confirmed mass of PSR J0348+0432 is also shown. The space below this lower limit is excluded. Thus, the interaction curve of this plane with the causality surface sets a boundary in the $J_{\rm{sym}}$ versus $K_{\rm{sym}}$ plane.  This boundary is the same as the solid line in the lower window of Fig. \ref{cau1}.
As we shall discuss later, this boundary sets the lower limit of nuclear symmetry energy. Again, it is clearly seen that the causality surface sets an absolutely upper limit for the mass of NSs at M$_{\rm{max}}=2.4$ M$_{\odot}$ almost independent of the EOSs used. This finding has some interesting implications. Given the model-independent nature of inverting the TOV for necessary EOSs to reproduce an observable and the general requirement of causality, the predicted absolutely maximum mass of 2.4 M$_{\odot}$ is rather general. While the allowed maximum mass varies between 2.01-2.4 M$_{\odot}$ depending on the high-density behavior of symmetry energy as indicated by the variation of the causality surface (blue) with $J_{\rm{sym}}$ and $K_{\rm{sym}}$.

The composite mass of the two NSs in GW170817 is 2.74 M$_{\odot}$ \cite{LIGO1}. The fate of the remanent in GW170817 is not observationally determined because of the
limited sensitivities of the current gravitational wave detectors \cite{LIGO1}. Thus the nature of the central remnant of GW170817 remains an open question. While it is generally proposed that the large composite mass of GW170817 would lead to a shortly lived hyper-massive NS or directly produce a black hole, there is no clear evidence to support or rule out a long-lived NS as the merger remnant, see, e.g. refs. \cite{Bau17,GW-Fate} for detailed discussions.
The causality surface clearly forbids the formation of a permanent NS as massive as 2.74 M$_{\odot}$. Indeed, several recent analyses of GW170817 data and numerical general relativity simulations
have placed upper bounds on the remanent mass in the range of $(2.15-2.32)$ M$_\odot$ \cite{Bau17,Margalit17,Shibata17,Rezzolla18,Ruiz18,Radice18,Ang-mass}
if a long-lived super-massive NS can be formed in the aftermath of GW170817. These findings are consistent with the indications of our Fig. \ref{mass1}.

\begin{figure}
\begin{center}
  \resizebox{0.7\textwidth}{!}{
 \includegraphics{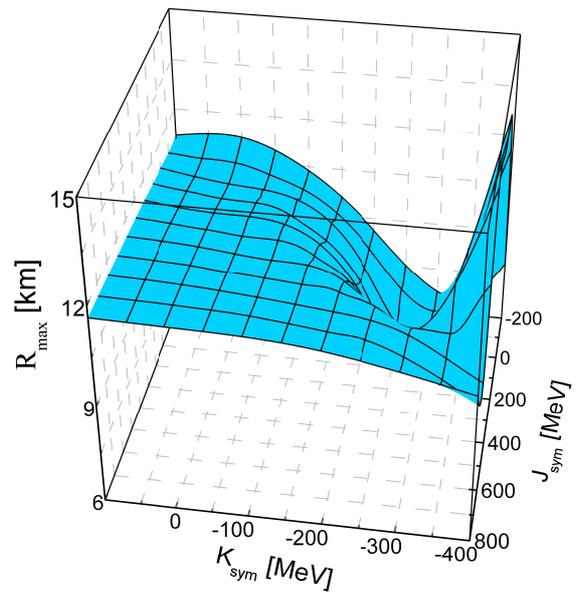}
}
\end{center}
  \caption{(color online) The radius of the most massive neutron stars on the causality surface as functions of
  $J_{\rm{sym}}$ and $K_{\rm{sym}}$, respectively. }
\label{Radius1}       
\end{figure}

Effects of the symmetry energy $E_{\rm{sym}}(\rho_0)$ and its slope $L(\rho_0)$ at saturation density $\rho_0$ on the radii of NSs have been studied extensively in the literature, see, e.g. ref. \cite{Prakash88,LiSteiner,Xu09} for earlier examples and ref. \cite{Lattimer14} for a recent review. However, less is known about effects of the high-density symmetry energy on the radii of NSs. It is thus interesting to study how the radius of the most massive NS on the causality surface evolves with $K_{\rm{sym}}$ and $J_{\rm{sym}}$. Indeed, as shown in Fig. \ref{Radius1}, these parameters characterizing the high-density symmetry energy have significant effects on the radii of the most massive NSs. It is seen that for $K_{\rm{sym}}$ higher than about $-100$ MeV regardless of the $J_{\rm{sym}}$ values the radius $R_{\rm{max}}$ stays at a constant of about 12 km. As the $K_{\rm{sym}}$ decreases especially with negative $J_{\rm{sym}}$ values, the symmetry energy becomes softer, the NS matter at $\beta$ equilibrium becomes more neutron rich. Both the mass and radius become smaller. In fact, the maximum mass drops below 2.01 M$_{\odot}$ as shown in the left window of Fig. \ref{mass1}. While it is in a region of parameter space already excluded due to its inability to support NSs with the minimum maximum masse of 2.01 M$_{\odot}$, it is worth noting that at the far corner of very negative $K_{\rm{sym}}$ and $J_{\rm{sym}}$ values where the symmetry energy becomes zero or negative, the very low-mass NSs containing almost pure neutron matter can be stable with large radii indicated by the raising $R_{\rm{max}}$.
%
\begin{figure}
\begin{center}
\resizebox{0.5\textwidth}{!}{
 \includegraphics{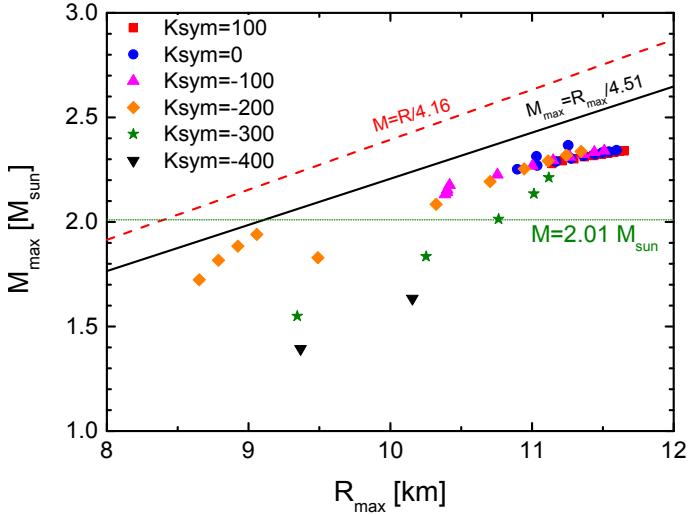}
}
\end{center}
  \caption{(color online) Correlation between the mass and radius of the most massive neutron stars on the causality surface in comparisons with the two scalings indicated by the solid and dashed lines.
  The minimum maximum mass of 2.01 M$_{\odot}$ is shown as a reference.}
\label{cau2}       
\end{figure}

\begin{figure*}
\begin{center}
\resizebox{0.45\textwidth}{6cm}{
 \includegraphics{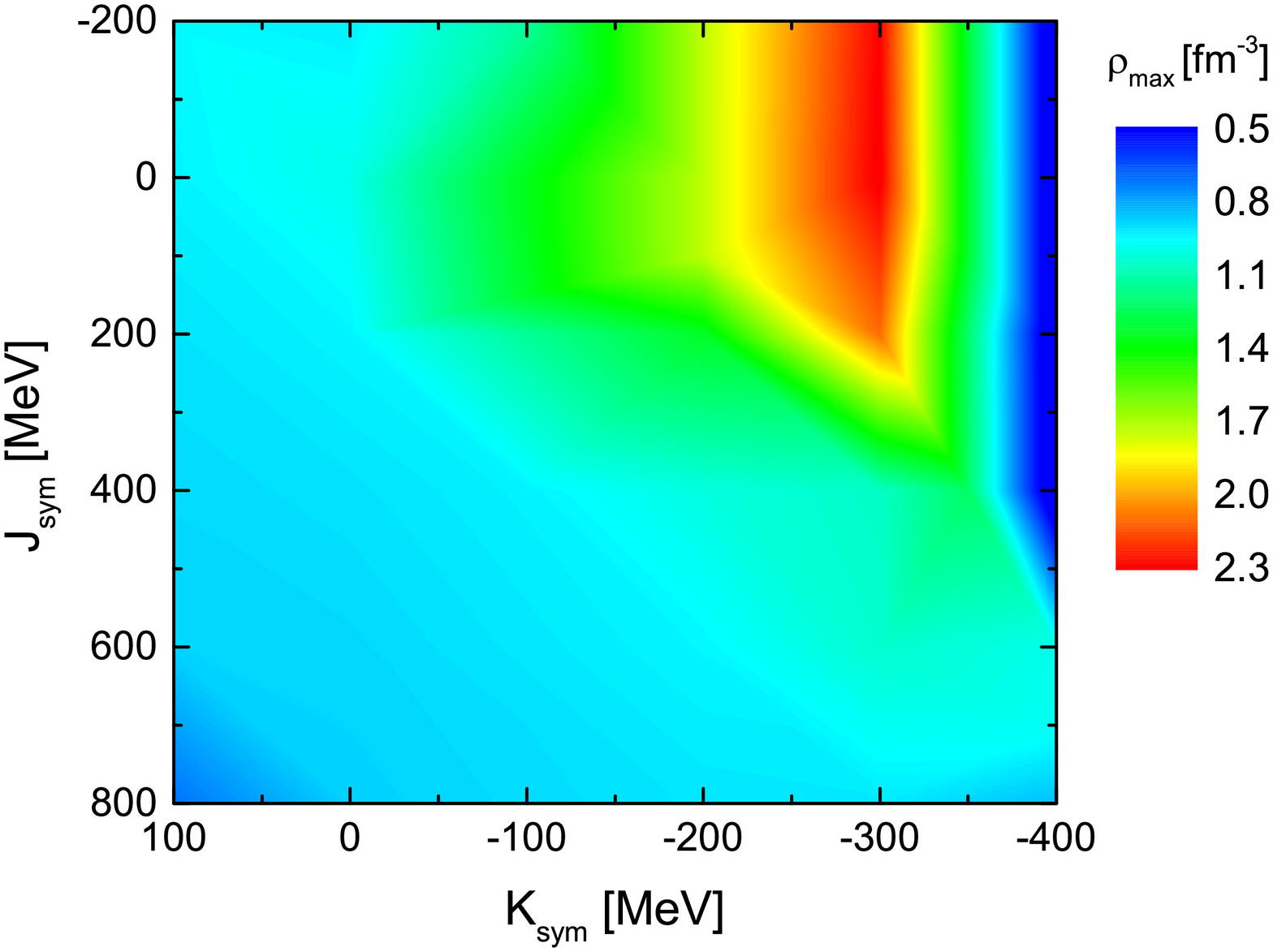}
 }
 \resizebox{0.49\textwidth}{6cm}{
  \includegraphics{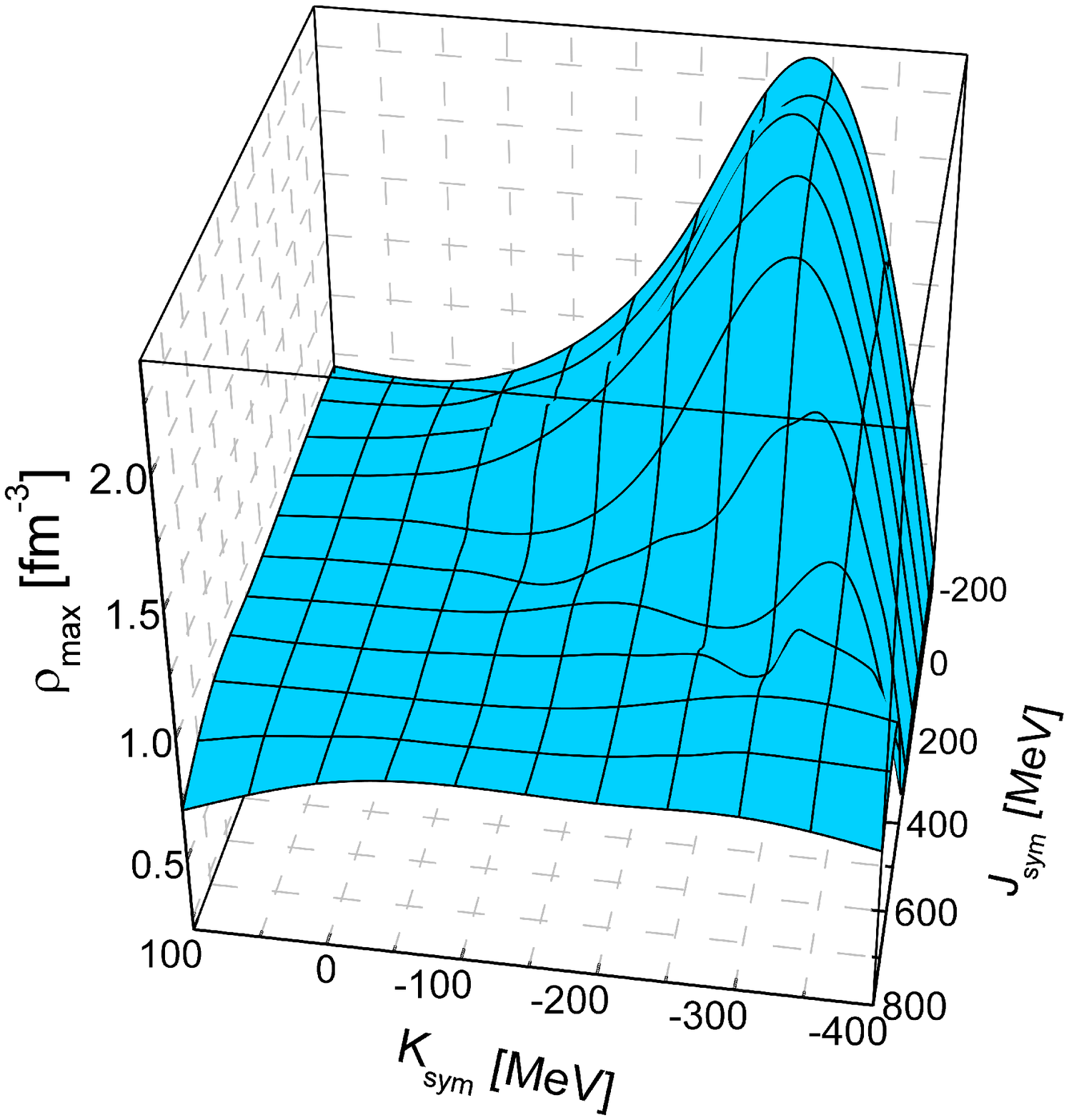}
}
\end{center}
  \caption{(color online) Central baryon density of the most massive neutron stars on the causality surface in the
  $J_{\rm{sym}}$ versus $K_{\rm{sym}}$ plane presented as contours (left) and functions in 3D (right), respectively.}
\label{Den}       
\end{figure*}

\subsection{Mass-radius correlation on the causality surface}
To this end, it is necessary and important to compare our M$_{\rm{max}}$ versus $R_{\rm{max}}$ relationship on the causality surface with the causality constraints on the M-R relationship widely used in the literature.
 An excellent review by Lattimer and Parakash on this topic can be found in ref. \cite{Latt07} and its update in ref. \cite{JLM11}. The most frequently used causality constraint is the one derived from the finding by Lindblom (1984) that the redshift has a maximum almost independent of the only parameter $\rho_f$ in his EOS for NS matter \cite{Lin84}.
For comparisons, it is important to first recall briefly how his EOS was constructed to respect the causal limit. In Lindblom's work and several subsequent studies confirming his results (for a complete list see refs. \cite{Latt07,JLM11}), they constructed EOSs that all have $v^2_s=c^2$ above a fiducial transition-density $\rho_f$, namely,  the EOS has the simple form $P(\epsilon)=\epsilon-\epsilon_f+P_f(\epsilon_f)$, for $\epsilon\ge\epsilon_f(\rho_f)$.
Below the $\rho_f$ considered to be in the envelopes of NSs, either some empirical or ``realistic" nuclear EOSs, such as the BPS EOS, or a
pure neutron matter EOS were used. The $\rho_f$ was taken as a model parameter. The redshift was shown to only weakly depend on the value of $\rho_f$.
For $\rho_f\ge3\times10^{14}$g cm$^{-3}$ (note that $\rho_0=2.7\times10^{14}$ g cm$^{-3}$ corresponding to a baryon density of 0.16 fm$^{3}$), the redshift was found to be
\begin{equation}
z={1\over\sqrt{1-2GM/Rc^2}}-1\le0.863,
\label{z}
\end{equation}
leading to a limit on the mass-radius relationship
\begin{equation}
GM\le Rc^2/2.83.
\label{rm1}
\end{equation}
In unit of M$_{\odot}$ for the mass and km for the radius, the above relation can be rewritten as
\begin{equation}
\label{rm}
\frac{M}{M_{\odot}} \le \frac{R}{{\rm km}}/4.16.
\end{equation}

The constraint of Eq. (\ref{rm}) has been widely used in the literature as a very general upper causal limit for the stiffness of all EOSs of NS matter. We notice that it was also emphasized by Lindblom that because the EOS reached the causal (stiffness) limit already at $\rho_f$ which can be quite low, the maximum mass supported can be rather high. Quantitatively, for the $\rho_f$ changing between 1 to 5 times $10^{14}$ g cm$^{-3}$, the maximum mass M$_{\rm{max}}$ varies between 6.71-3.02 M$_\odot$ far above the absolutely maximum mass we discussed above, while the redshift z varies only in a very small range between 0.883-0.858 \cite{Lin84}, respectively.
According to ref. \cite{JLM11}, the Eq. (\ref{rm1}) is essentially the same result found earlier by Rhoades and Ruffini in 1974 (although they don't mention it in their paper, it is implicit in their formulation) \cite{RR74}.  The same result was found by a different method by Glendenning in 1992 \cite{Glen92}. The Rhoades and Ruffini result was later made explicit with the methodology of Koranda, Stergioulas and Friedman (1997) \cite{Kor97} and elaborated by Lattimer and Prakash in 2011 -- the so-called maximum compactness causal limit \cite{JLM11}. 

Instead of allowing $v^2_s=c^2$ in all regions at densities higher than the $\rho_f$ in all NSs regardless of their masses, on our causality surfaces the condition $v^2_s=c^2$ is satisfied only at the central density in the most massive NS configuration in the M-R sequence calculated for a given EOS. All other NSs in the sequence still have $v_s$ less than c in all density regions reached. Thus, the maximum masses M$_{\rm{max}}$ on our causality surfaces are expected to be below the ones inferred from Lindblom's causality constraint of Eq. (\ref{rm1}) or Eq. (\ref{rm}). This is demonstrated numerically in Fig. \ref{cau2} by comparing the red dashed line for M=R/4.16 from Eq. (\ref{rm}) with our results (scattered symbols). The minimum maximum mass of M$_{\rm{max}}=2.01$ M$_{\odot}$ is also shown as a reference.
The values of  M$_{\rm{max}}$ and $R_{\rm{max}}$ are taken from Fig. \ref{mass1} at the same $K_{\rm{sym}}$ and $J_{\rm{sym}}$ coordinates on the lattice grids. For example, for $K_{\rm{sym}}=100$ MeV with $J_{\rm{sym}}$=-200, -100, 0,..., 800 MeV, both the M$_{\rm{max}}$ and $R_{\rm max}$ are almost constants, the resulting M$_{\rm{max}}-R_{\rm max}$ correlation, shown as red squares near the right frame, is almost a constant of
M$_{\rm{max}}=2.4 $ M$_{\odot}$ independent of the symmetry energy parameters used. This is because near the absolutely maximum mass, the pressure is dominated by the EOS of SNM instead of the symmetry energy.  As the $K_{\rm{sym}}$ and $J_{\rm{sym}}$ decrease to more negative values, making the symmetry energy softer, the M$_{\rm{max}}$ decreases slower than the radius $R_{\rm max}$.
Comparing our results with the M-R correlation of Eq. (\ref{rm}), it is seen that near the limit of M$_{\rm{max}}=2.4$ M$_\odot$ and $R_{\rm max}=11.5$ km, our M$_{\rm{max}}$ values are about 10\% below the Eq. (\ref{rm}) prediction qualitatively consistent with the expectations discussed above. 

It is important to emphasize that the mass and radius from Lindblom's redshift limit of Eq. (\ref{rm1}) or Eq. (\ref{rm}) are not for the most massive NS configuration that his EOS can support. In fact, as we mentioned earlier, his EOS can support much more massive NSs as heavy as 6.71 M$_\odot$. It is thus also very interesting to compare directly with the maximum mass M$_{\rm{max}}$ and the corresponding radius $R_{\rm max}$ from earlier studies. It was shown in Eq. (10) of ref. \cite{Lat90} that for very soft SNM EOSs with an incompressibility $K_0<150 $ MeV and a fiducial transition-density $\rho_f\le 6\rho_0$ both the mass and radius of the most massive NSs scale approximately with $(\rho_0/\rho_f)^{1/2}$ according to 
\begin{equation}\label{lapp}
\frac{M_{\rm max}}{M_{\odot}}\simeq 4.1 \left(\frac{\rho_0}{\rho_f}\right)^{1/2};~\frac{R_{\rm max}}{{\rm km}}\simeq 18.5 \left(\frac{\rho_0}{\rho_f}\right)^{1/2}.
\end{equation}
Taking the ratio of the above expressions leads to the following  M$_{\rm max}$-R$_{\rm max}$ scaling independent of the fiducial transition-density $\rho_f$
\begin{equation}
\label{MR-S}
\frac{M_{\rm{max}}}{M_{\odot}} \approx \frac{R_{\rm max}}{{\rm km}}/4.51.
\end{equation}
It is slightly lower than that of Eq. (\ref{rm}) probably because the smaller value of $K_0$ used in deriving it. 
The approximations in Eq. (\ref{lapp}) were obtained by using similarly constructed EOSs as in deriving the scaling of Eq. (\ref{rm}) assuming $v^2_s=c^2$ above the fiducial transition-density $\rho_f$. It is important to note that the Eq. (\ref{MR-S}) is from actually estimating the M$_{\rm{max}}$ and $R_{\rm max}$ for the most massive NSs supported by the EOS used instead of using the approximate invariance of the redshift in deriving the Eq. (\ref{rm}) for all NSs.  It is seen in Fig. \ref{cau2} that this scaling is closer than that of Eq. (\ref{rm}) to our results. More quantitatively, near the limit of M$_{\rm{max}}=2.4$ M$_\odot$ and $R_{\rm max}=11.5$ km, our M$_{\rm{max}}$ values are about 5\% below the Eq. (\ref{MR-S}) prediction. The 5\% to 25\% differences between our numerical results and predictions of the two scalings of Eqs. (\ref{rm} and \ref{MR-S}) quantify the model dependence in estimating the causality constraint on the mass-radius relation. Given our limited knowledge about NSs, such level of model dependence is not so serious.

We notice that the $R_{\rm max}$ from our study is between 10.5 and 11.5 km above the minimum maximum mass of M$_{\rm{max}}=2.01$ M$_{\odot}$. This sets a lower limit for the radii of most massive NSs, which is consistent with existing observations as we shall discuss later. The predicted NS maximum masses using various EOSs in the literature still spread out in a large range. There are indeed models predicting maximum masses higher than 2.4 M$_{\odot}$ but below the scaling of Eq. (\ref{rm}), see, e.g. examples given in refs. \cite{GW-Fate,Ang-quark,Ko18}. However, we emphasize here again that the absolutely maximum mass of M$_{\rm{max}}$=2.4 M$_{\odot}$ found here is based on a minimum NS model. It does not rely on any prior knowledge about the EOS but the physics requirement that the causal limit is reached only at central densities of the most massive NSs supported by the EOSs considered. The absolutely maximum mass is the maximum of the maximum masses supported by all EOSs while the causality is respected. It is therefore theoretically EOS independent. This is very different from comparing the predicted maximum masses using various EOSs with an estimated universal M-R relationship from general causality considerations under some assumptions about the EOS in NSs. Thus, those EOSs predicting the maximum masses higher than 2.4 M$_{\odot}$ but still below the estimated M-R causality line are not guaranteed to be actually causal as the sound speed $v_s$ depends on the specific EOS used in the model.

\subsection{Causality limit on the maximum density reachable in neutron stars and effects of high-density nuclear symmetry energy}
Besides the discussions above, one important question also needs to be answered.
What are the ultimate baryon and energy densities of observable cold matter? Lattimer and Prakash pointed out that the largest observed mass of NSs sets the lowest upper limit on the maximum energy density \cite{Lat-Pra}. Theoretically, it is interesting to known not only how high but also where and under what conditions the highest densities are reached in NSs.  Answers to these questions have several profound implications \cite{Lat-Pra}.
Within our minimum model for NSs, the central density reached in the most massive NSs on the causality surface provides useful hints to answer these questions.

The central baryon density $\rho_{\rm max}$ of the most massive NSs on the causality surface is presented as contours in the plane of $K_{\rm{sym}}$ versus $J_{\rm{sym}}$ in the left window of Fig. \ref{Den}. While in the right window, it is presented as a function in 3D. From both graphs, it is seen consistently that the $\rho_{\rm max}$ is about a constant of $5\rho_0$ in the lower left corner where the $K_{\rm{sym}}$ and $J_{\rm{sym}}$ are both positive, leading to stiff symmetry energies. However, it increases quickly as the $K_{\rm{sym}}$ and $J_{\rm{sym}}$ move toward their lower limits near the upper right corner where the symmetry energy becomes super-soft, supporting only light and very neutron-rich NSs.

Since the radius decreases much faster than the mass as the symmetry energy changes from being stiff to super-soft and the density scales with $M/R^3$, the $\rho_{\rm max}$ thus varies very dramatically with $K_{\rm{sym}}$ and $J_{\rm{sym}}$ as they become negative, towards super-soft symmetry energies.
The $\rho_{\rm max}$ reaches as high as $15\rho_0$. However the peak is in the region excluded by the requirement that all EOSs have to support at least the minimum maximum mass of 2.01 M$_{\odot}$ as illustrated in Fig. \ref{mass1} already. Consequently, the observed most massive NS limits the $\rho_{\rm max}$ to the range of 5-9 $\rho_0$ depending on the high-density symmetry energy. 

Of course, as for all minimum models for NSs, there is a maximum range beyond which new physics ingredients, such as non-nucleonic degrees of freedom, various new phases and possible boson condensations, have to be added. The $\rho_{\rm max}$ is an indicator when such new physics should be incorporated. The rather high $\rho_{\rm max}$ revealed in Fig. \ref{Den} reminds us that our results obtained within the minimum model have to be interpreted with some caveats.

\begin{figure}
\begin{center}
\resizebox{0.45\textwidth}{!}{
 \includegraphics{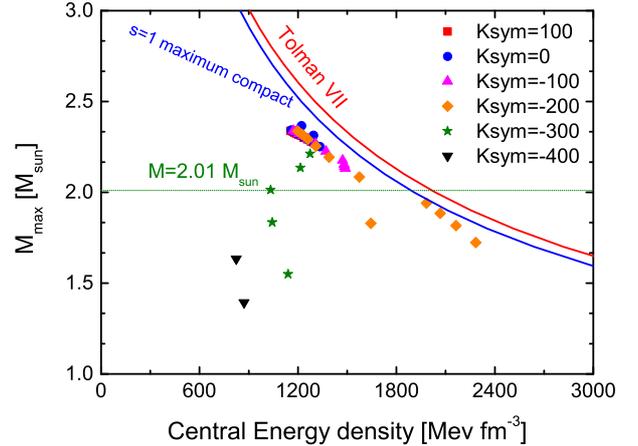}
}
\end{center}
  \caption{(color online) The masses of the most massive neutron stars on the causality surface as a function of their central energy density. The estimates using the Tolman VII solution of Einstein's equations from ref. \cite{Lat-Pra} and the ansatz of most compact neutron stars of ref. \cite{Kor97} are shown with red and blue lines, respectively.}
\label{Den-M}       
\end{figure}
To see the maximum energy density reached in most massive NSs,  shown in Fig. \ref{Den-M} are the masses versus central energy densities for the most massive NSs on the causality surface.
The labels are the same as we used in Fig. \ref{cau2}. It is useful to know that for M$_{\rm{max}}\geq 2.01$ M$_{\odot}$, the maximum energy density is between about 1100 to 1700 MeV/fm$^{-3}$. For the absolutely maximum mass NS with 2.4 M$_{\odot}$, the central energy density is about 1200 MeV/fm$^{-3}$ independent of the symmetry energy parameters. As the M$_{\rm{max}}$ decreases to significantly below 2.01 M$_{\odot}$, effects of the symmetry energy parameters become larger mainly through their influences on the radii as we explained earlier.

It is interesting to compare our numerical results with two analytical approximations of the maximum energy density reachable in NSs.
In ref.  \cite{Lat-Pra}, Lattimer and Prakash estimated the maximum energy density using the Tolman VII  analytical solution of Einstein's field equations.
It was shown that all analytical solutions of Einstein's field equations for fixed compactness obey the scaling $\epsilon_{\rm{max}} \propto 1/M_{\rm{max}}^2$.  Moreover, the Tolman VII solution gives quantitatively
the upper limit \cite{Lat-Pra}
\begin{equation}
\epsilon_{\rm{max}}\cdot M_{\rm{max}}^2=1.53\times10^{16}{\rm~M}_\odot^2 {\rm~g~cm}^{-3}.
\end{equation}
The above scaling is shown in Fig. \ref{Den-M} with the red line.
Another estimate considered as more fundamental \cite{JLM11} is based solely on causality by constructing an idealized EOS for the most compact NS by Koranda, Stergioulas and
Friedman \cite{Kor97}. In such stars, the pressure is set to zero (thus most soft) below a cut-off energy density $\epsilon_0$ and has the form $p=s(\epsilon-\epsilon_0)$ (most stiff) above $\epsilon_0$ where $s=v^2_s$ is a constant representing the square of the adiabatic sound speed. Setting $v^2_s=c^2$, the energy density of the most compact NS
when expressed in astrophysical units satisfies the relation \cite{JLM11}
\begin{equation}
\epsilon_{\rm{max}}\cdot M^2_{\rm{max}}= 1.36 \times 10^{16} {\rm~M}_\odot^2 {\rm~g~cm}^{-3}.
\end{equation}
This upper bound is shown with the blue line in Fig. \ref{Den-M}. The above scaling is consist with the Eq. \ref{MR-S} \cite{JLM11}.
The above two scalings are in parallel and differ by only a few percent. 

Our results for NSs above the minimum maximum mass of 2.01 M$_{\odot}$ approximately follow the two scalings. As one expects, our results are very close but fall below the above two scalings. This is because in our method the EOS may becomes acausal only at the central density of the most massive NS, instead of assuming $v^2_s=c^2$ at {\it all} densities above the assumed cut-off energy density $\epsilon_0$. Moreover, in our approach the $\epsilon_0$ is being searched and possibly found by checking if $v^2_{s}=c^2$ can ever be reached in the most massive NS for a given set of EOS parameters, without pre-assuming its existence.
The observed differences between our results and the above two scalings as well as other estimates in the literature indicates again the model dependence in predicting the absolutely maximum mass of NSs. While we can not prove
our prediction is model independent and we do not claim our prediction is more accurate than any other estimates, it is useful to emphasize that the absolutely maximum mass of 2.4M$_{\odot}$ found here is almost completely independent of the EOSs used, indicating its likely model independence. Nevertheless, it would be interesting to study in the future if this absolutely maximum mass is universal in a really model independent way. We are making efforts in this direction.

In summary of this section, neutron stars' absolutely maximum mass is found to be 2.4 M$_{\odot}$ from investigating the causality surface. It is approximately independent of the EOSs used. Depending on the high-density behavior of nuclear symmetry energy, the M$_{\rm{max}}$ ranges between 2.01-2.4 M$_{\odot}$. The causality surface together with the mass of PSR J0348+0432 set a lower boundary in the plane of $K_{\rm{sym}}$ versus $J_{\rm{sym}}$, limiting the high-density behavior of nuclear symmetry energy. They also limit the radii and maximum density reachable in NSs. These causality limits will be respected and used in our following analyses.
\begin{figure*}
\begin{center}
\resizebox{0.48\textwidth}{!}{
  \includegraphics{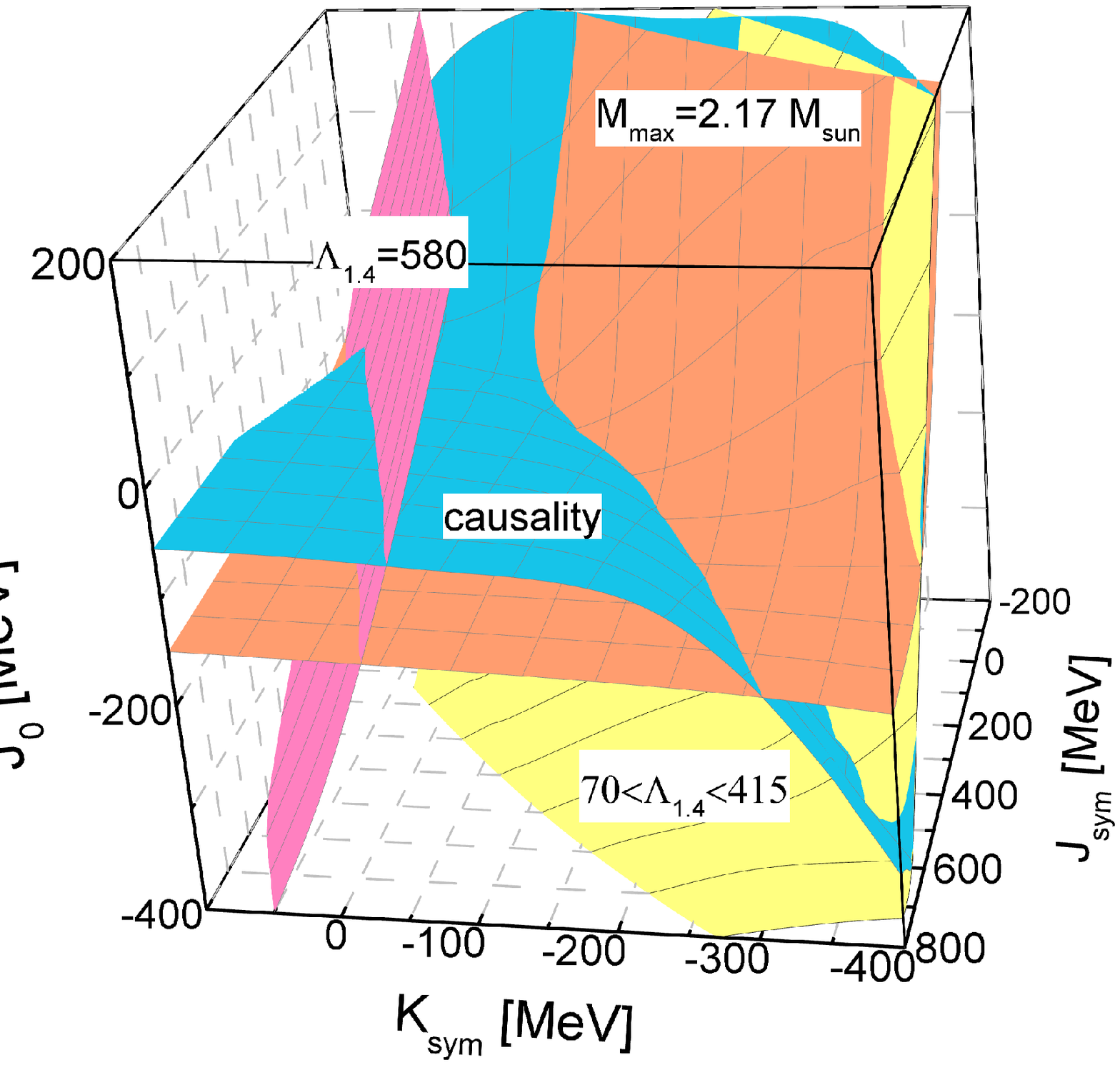}
  }
  \resizebox{0.48\textwidth}{!}{
  \includegraphics{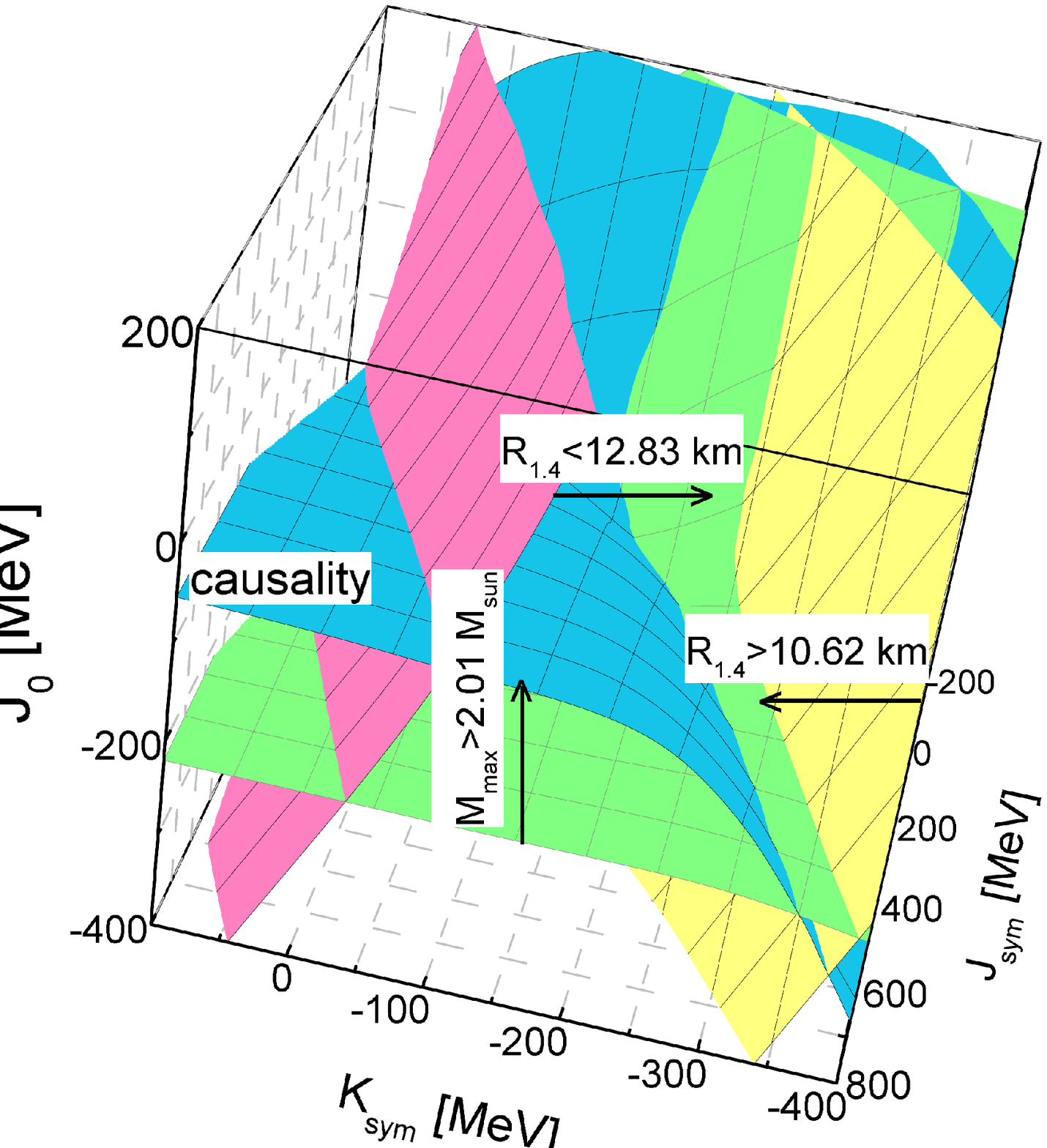}
}
\end{center}
  \caption{(color online) Observational restrictions on the 3 dimensional EOS parameter space in $K_{\rm{sym}}$, $J_{\rm{sym}}$ and $J_0$. Left: the magenta and yellow surfaces have the tidal deformability for
 canonical neutron stars $\Lambda_{\rm{1.4}}=580$ and $70\leq \Lambda_{\rm{1.4}}\leq 415$, respectively. On the light-brown surface, the maximum mass of neutron stars is set at M$_{\rm{max}}=2.17$ M$_\odot$.
 Right: the green, magenta, yellow and blue surfaces represent M$_{\rm{max}}=2.01$ M$_{\odot}$, $R_{1.4}=12.83$ km, $R_{1.4}=10.62$ km and the causality surface, respectively.}
\label{obsers}       
\end{figure*}

\section{Observational restrictions on the parameter space of high-density nuclear EOS and symmetry energy}\label{obs}
Since the pioneering works of Lindblom \cite{Lin92} as well as Lattimer and Prakash et al. \cite{Lattimer01}, extensive studies by many people have confirmed that {\it important constraints on the EOS can be obtained with even a single radius measurement, if it is accurate enough, and that the quality of the constraint is not very sensitive to the mass \cite{Lattimer01}. While it is necessary to have a series of mass and radius measurements to accurately
constrain completely the dense matter EOS \cite{Lin92}.} Significant efforts and much progress have been made in measuring/calculating the radii of NSs using various probes/models while some issues/uncertainties remain to be resolved/reduced. In this section, using widely accepted results of mass and radius measurements available in the literature, we examine how they may restrict the 3D EOS parameter space in $K_{\rm{sym}}$, $J_{\rm{sym}}$ and $J_0$. Moreover, we shall obtain crosslines of surfaces representing constant radii and the minimum maximum mass of NSs as well as the causal limit.  These crosslines will be used in the next section to set boundaries for the high-density EOS and symmetry energy.
\begin{figure*}
\begin{center}
\resizebox{0.9\textwidth}{!}{
  \includegraphics{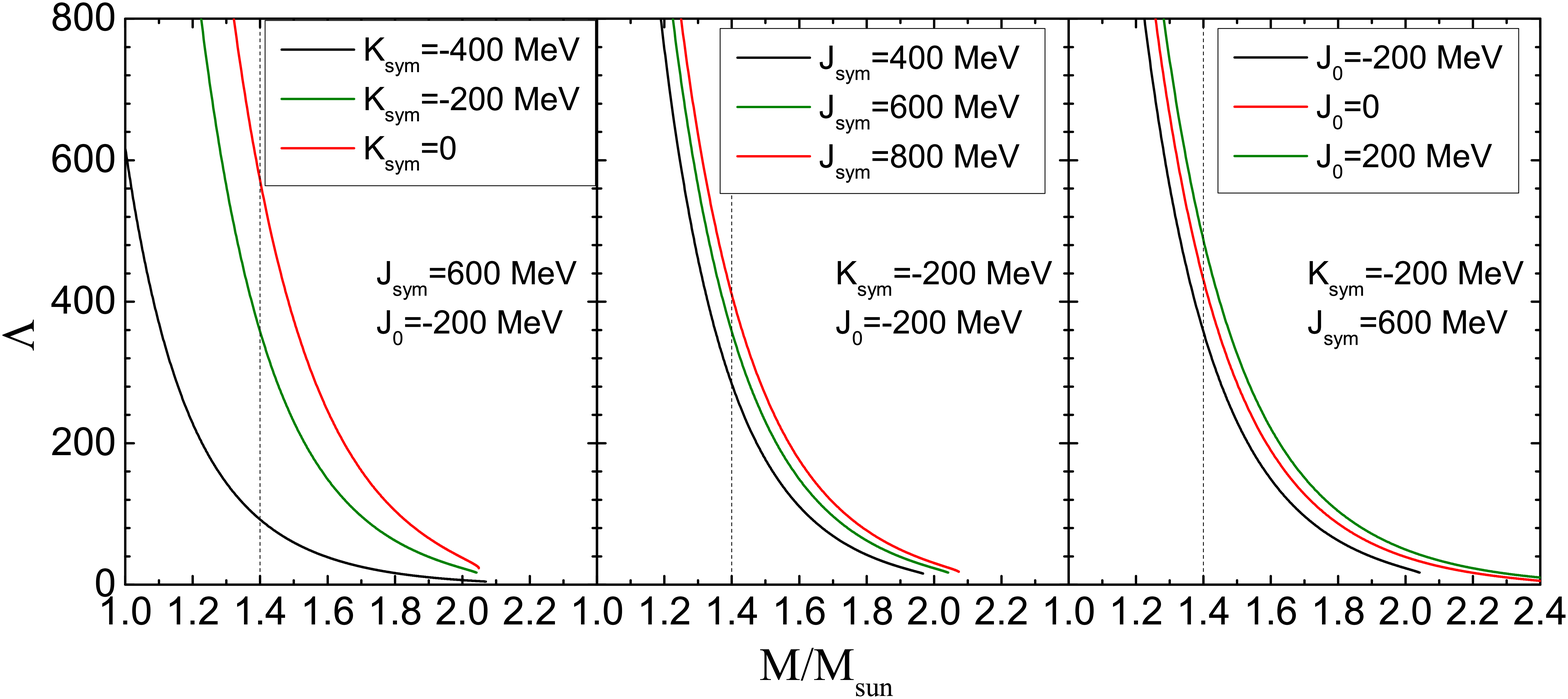}
}
\end{center}
  \caption{(color online) Tidal deformability as a function of mass of neutron stars by varying the high-density EOS parameter $K_{\rm{sym}}$, $J_{\rm{sym}}$  and $J_0$, respectivity.
}
\label{LM}       
\end{figure*}

\subsection{What we know about the radii of neutron stars}
Let us first summarize what we have learned about the radii of NSs before trying to use them to constrain the high-density EOS and symmetry energy.
The radii of NSs have been measured using several approaches. For example, the radii of 14 NSs have been extracted from analyzing thermal emissions from quiescent low-mass X-ray binaries (QLMXBs) and photospheric radius expansion (PRE) bursts \cite{Steiner10,Ozel16,Bog16,Raithel16,Raithel17,Steiner17}.  While there are still some variations and issues in different analyses of the same data, see e.g., ref. \cite{CMiller} for a recent review, the radius $R_{\rm{1.4}}$ of canonical $1.4$ M$_\odot$ NSs was found generally to be around $10.62\leq R_{\rm{1.4}}\leq12.83$ km \cite{Lattimer14}. The upper limit of the dimensionless tidal deformation $\Lambda_{\rm{1.4}}\leq 800$ for canonical NSs from the first analysis of GW170817 by the LIGO+Virgo Collaborations \cite{LIGO1} has been used in a number of studies to extract the $R_{\rm{1.4}}$. While the lower limit has somewhat larger variations in different studies, a rather consistent upper limit of about $R_{\rm{1.4}}\leq 13.7$ km has been found \cite{Ang-mass,Ann,Fattoyev17,Most,Plamen3,Rai,Tew18,Malik,Holt18}. An improved analysis of the GW170817 event by the LIGO+Virgo Collaborations has now provided both the upper and lower limits for the tidal deformability. The $\Lambda_{\rm{1.4}}$ is found to be in the range of $70-580$ \cite{LIGO2018}, leading to a refined constraint on the radius in the range of $R_{\rm{1.4}}$= 10.5-13.3 km. While an independent reanalysis of the GW170817 data found that the $\Lambda_{\rm{1.4}}$ depends on the mass priors used in their Bayesian analyses, a common radius of $8.9\leq R_{\rm{1.4}}\leq13.2$ km across all mass priors was reported \cite{De2018} largely in agreement with that found in the new analyses by the LIGO+Virgo Collaborations \cite{LIGO2018}. It is also very interesting to mention that $R_{\rm{1.4}}\leq 14.4$ km was obtained earlier from studying the fastest-spinning radio pulsar ever found and confirmed, the PSR-J1748-2446ad spinning at 716 Hz \cite{Hes06}.

To this end, we emphasize since it is often overlooked that terrestrial nuclear experiments, especially heavy-ion reactions with radioactive beams at intermediate energies, do provide strong constraints on the EOS of dense neutron-rich matter and subsequently the radii of NSs as reliably as astrophysical observations mentioned above, see, e.g. refs. \cite{Dan02,Li08} for earlier reviews and ref. \cite{Tsang2018} for an example of very recent studies. As an earlier example from 2006, using the MDI (Momentum-Dependent Interaction) energy density functional \cite{Das03,Lwchen05,LiBA05} with its EOS for SNM constrained up to about $4.5\rho_0$ by nucleon collective flow and kaon production data \cite{Dan02} in relativistic heavy-ion collisions, and its symmetry energy term constrained up to about $1.2\rho_0$ by isospin diffusion data \cite{Tsang04} in intermediate energy heavy-ion collisions then extrapolated to higher densities based on the MDI energy density functional, the radius of canonical NSs was predicted to be 11.5 km $\leq R_{\rm{1.4}} \leq 13.6$ km~\cite{LiSteiner}. The same MDI EOS was also used to place bounds on the
quadrupole deformation of isolated neutron stars, strain amplitude as well as the frequency and damping time of several modes of NS oscillations which are all potential sources of gravitational waves \cite{Plamen1,Plamen2,DeHua,Newton}. Ironically, given all the existing uncertainties in every method used to extract the radii of NSs, the predicted radius range using the MDI EOS partially constrained by the terrestrial nuclear reaction experiments is in very good agreement with the astrophysical observations mentioned above.

Overall, it is very encouraging to see that the constraints on the radius $R_{\rm{1.4}}$ especially its upper limit from all analyses are rather consistent. This impressive consistency further validates and signifies the multi-messengers approach of constraining the EOS of dense neutron-rich matter using all available probes in both astrophysical observations and terrestrial experiments. In the following, we adopt the range $10.62 \leq R_{\rm{1.4}}\leq 12.83$ km from analyzing X-rays \cite{Lattimer14} as an example in constraining the high-density EOS parameters.  This range is almost identical to that of $10.42 \leq R_{\rm{1.4}}\leq 12.80$ km extracted from a very recent analysis of the tidal deformability in GW170817 within a Bayesian statistical framework incorporating constraints on the EOS from laboratory measurements of nuclei and state-of-the-art chiral effective field theory methods \cite{Holt18}.

Since the relationship between the tidal deformability $\Lambda$ and radius $R$ is EOS-model dependent \cite{NBZ-JPG} and the current uncertainty of the measured $\Lambda$ from the single event GW170817 is still relatively larger than those in measuring the radius $R$ using some other approaches, we study separately in the left and right blocks of Fig. \ref{obsers} how the existing results of $\Lambda_{\rm{1.4}}$ and $R_{\rm{1.4}}$ together with the mass measurements and causality condition can limit the 3D EOS parameters. For a given observable, we perform an inversion of the TOV equation by looping through the 3D EOS parameters. While a single observable can often restrict the parameter space, large degeneracies in the EOS parameters are expected. In fact, a constant surface of a given observable can only give the required combinations of the EOS parameters to reproduce the observed value. The crosslines of two constant surfaces reduce the degeneracies. To uniquely determine all three parameters would require at least three surfaces to cross at a point. This is consistent with the earlier expectation that the measurements of masses and radii of 2-3 NSs are necessary to completely determine the EOS of NS matter \cite{Lin92}.

\subsection{What we learn about the EOS from the tidal deformability of neutron stars}
In the left block of Fig. \ref{obsers}, the magenta surface of a constant $\Lambda_{\rm{1.4}}=580$ is the upper limit of tidal deformability from the improved analysis of GW170817
by the LIGO+Virgo Collaborations. It is rather vertical and covers the whole ranges of both $J_0$ and $J_{\rm{sym}}$ considered, but spans a narrow range in $K_{\rm{sym}}$.
Since the $J_0$ and $J_{\rm{sym}}$ are coefficients of the $(\rho/\rho_0-1)^3$ terms in our parameterizations of the EOS, it is easy to understand that the $\Lambda_{\rm{1.4}}$ is not sensitive to really high-density EOS as the average density reached in $1.4$ M$_{\odot}$ NSs are considered intermediate that is not high enough for the $J_0$ and $J_{\rm{sym}}$ to play really big roles. On the other hand, the $K_{\rm{sym}}$ as the coefficient of the $(\rho/\rho_0-1)^2$ term in parameterizing the symmetry energy plays a bigger role at the intermediate densities reached in canonical NSs. This means that the upper limit of $\Lambda_{\rm{1.4}}$ is insensitive to the EOS of SNM but can limit the symmetry energy to a narrow region through its constraints on the $K_{\rm{sym}}$ parameter. Very low values of $J_0$ and/or $K_{\rm{sym}}$ as well as $J_{\rm{sym}}$ together make the EOSs too soft to support NSs with masses as high as $1.4$ M$_{\odot}$. In fact, only regions with large $J_0$ can reach the lower limit of $\Lambda_{\rm{1.4}}=70$ if the symmetry energy is very soft. To make our presentations clear while still convey the main physics, we use the yellow surface to indicate the integrated parameter space leading to $70\leq \Lambda_{\rm{1.4}}\leq 415$.
We note that the crossline between the yellow surface and the bottom surface is not corresponding to $\Lambda=415$. In fact, only one point at ($K_{\rm{sym}}=35$ MeV, $J_{\rm{sym}}=147$ MeV and $J_0=-400$ MeV)
has $\Lambda=415$. This point is the nearest point to the surface with $\Lambda=580$.

While the 3D plots in Fig. \ref{obsers} are necessary for us to set the limits on the EOS parameters by examining the crosslines of different constant surfaces corresponding to various observations and physics constraints,
they are not easy to decipher.  To reveal the individual roles of the high-density EOS parameters on the tidal deformability, shown in Fig. \ref{LM} is the tidal deformability as a function of NS mass by varying the high-density EOS parameter $K_{\rm{sym}}$, $J_{\rm{sym}}$ and $J_0$, respectively.  By fixing two of the parameters in the relevant regions of the 3D EOS space, sensitivities of the $\Lambda$ on the three high-density EOS parameters can be clearly reviewed. Comparing the three windows of Fig. \ref{LM}, it is seen that the $\Lambda$ is most sensitive to the $K_{\rm{sym}}$ while effects of varying the $J_{\rm{sym}}$ and $J_0$ are clearly visible.
For canonical NSs of 1.4M$_{\odot}$, many combinations of the three high-density EOS parameters can lead to the same $\Lambda$.  To evaluate effects of the three parameters on equal footing and identify all combinations of the EOS parameters leading to the same observables, the 3D plots are necessary although they are sometimes difficult to be interpreted. 

The causality surface (blue) limits the $J_0$ from the above.  Interestingly, however, the $J_0$ may not be able to
go as high as the causality surface indicates. As we mentioned earlier, several studies have estimated the maximum mass of the possible super-massive remanent produced in the immediate aftermath of GW170817.
For example, using electromagnetic constraints on the remnant imposed by the kilonova observations and the gravitational wave information,  Margalit and Metzger \cite{Margalit17}
found a maximum mass of M$_{\rm{max}}\leq 2.17$ M$_\odot$ with 90\% confidence. The constant surface of this mass is shown in light brown. It is seen that the crossline of this surface with the causality surface provides a lower boundary from the right for the $K_{\rm{sym}}$-$J_{\rm{sym}}$ relation significantly tighter than that along the crossline of causality and the condition $70\leq \Lambda_{\rm{1.4}}\leq 415$. Moreover, if confirmed, the M$_{\rm{max}}\leq 2.17$ M$_\odot$ together with the minimum maximum mass
of 2.01 M$_\odot$ shown as the bottom surface in the right block will limit the $J_0$ to the range of about $-200\pm 25$ MeV.

\subsection{What we learn about the EOS from the radii of neutron stars}
In the right block of Fig. \ref{obsers}, the green, magenta and yellow surfaces represent M$_{\rm{max}}=2.01$ M$_{\odot}$, $R_{1.4}=12.83$ km and $R_{1.4}=10.62$ km, respectively. The allowed regions are indicated by the arrows. First of all, comparing the two magenta surfaces in the left and right blocks, it is seen that the one representing $\Lambda_{\rm{1.4}}=580$ is more vertically oriented than the one on the right representing $R_{1.4}=12.83$ km, indicating that a larger region in $K_{\rm{sym}}$ can be used to give the same radius compared to the one needed to give the same tidal deformability. While both surfaces span the whole ranges of $J_0$ and $J_{\rm{sym}}$ considered. This confirms the expectation that the tidal deformability is more sensitive to the symmetry energy than the radius itself and none of them is much affected by the high-density EOS of SNM. It is interesting to see that the causality and M$_{\rm{max}}=2.01$ M$_{\odot}$ surfaces together not only limit the range for $J_0$ from the top and bottom, respectively, their crossline also determines the $K_{\rm{sym}}$-$J_{\rm{sym}}$ relation along the right boundary. It is seen that the lower limit $R_{1.4}=10.62$ km of the radius is actually outside this boundary. Thus, if one believes in the causality and the measured minimum maximum mass of M$_{\rm{max}}=2.01$ M$_{\odot}$, the reported small radii of NSs less than about 10 km is clearly ruled out at least within the rather general model framework of this work. On the hand, it is seen that the crossline of the
M$_{\rm{max}}=2.01$ M$_{\odot}$ and $R_{1.4}=12.83$ km surfaces provide a limit on the $K_{\rm{sym}}$-$J_{\rm{sym}}$ relation from the left. The results shown in this block signifies the need of improving the
accuracy of radius measurements especially regarding the lower limit of $R_{1.4}$.

If we consider the tidal deformability from GW170817 and the radii from other measurements as independent and assume they are all equally accurate,
comparing the results in the two blocks of Fig. \ref{obsers}, it is seen that the radius constraint $R_{1.4}\leq 12.83$ km provides a slightly tighter constraint on
the $K_{\rm{sym}}$-$J_{\rm{sym}}$ boundary than the condition $\Lambda_{\rm{1.4}}\geq 580$. While the lower limits of both $R_{1.4}=10.62$ km and $\Lambda_{\rm{1.4}}\geq 70$ do not provide
any additional constraints as they are both outside the crossline between the causality surface and the minimum maximum mass constraint M$_{\rm{max}}\geq 2.01$ M$_{\odot}$. In the following, we thus use the
radius $R_{1.4}\leq 12.83$ km instead of  $\Lambda_{\rm{1.4}}\leq 580$ to set the left boundary in the $K_{\rm{sym}}$-$J_{\rm{sym}}$ plane. This choice is also physically necessary as we are interested in making an unbiased
comparison of the pressure extracted independently from our analyses directly using the radius data with that extracted from GW170817 by the LIGO+Virgo Collaborations. Using the $\Lambda_{\rm{1.4}}$ in our following analyses would introduce self-correlations. In addition, to ensure the thermodynamical stability through out the star,
we also use the condition that the transition pressure $P_t$ at the crust-core transition density is always positive as we discussed earlier. This latter condition has been discussed in detail in our earlier work in ref. \cite{NBZ2018a} where the crust-core transition point was found by explicitly investigating where the incompressibility of NS matter start to become negative.

\begin{figure}
\begin{center}
\resizebox{0.45\textwidth}{!}{
  \includegraphics{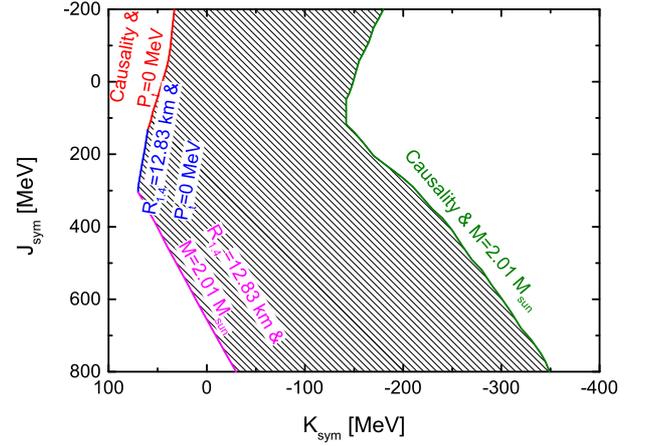}
}
\end{center}
  \caption{(color online) Boundaries of the allowed $K_{\rm{sym}}$-$J_{\rm{sym}}$ plane obtained from the crosslines shown in the right window of Fig. \ref{obsers}.}
\label{exc1}       
\end{figure}
Under the above conditions, the allowed region in the $K_{\rm{sym}}$-$J_{\rm{sym}}$ plane is shown in Fig. \ref{exc1} as the hatched area. Different segments obtained from the crosslines of two different observables and/or physical conditions are labled, separately. It is seen that the $J_{\rm{sym}}$ is still completely unconstrained with respect to its known uncertain range in the literature. This will lead to uncertainties in our extracted symmetry energy when the $(\rho/\rho_0-1)^3$ term becomes significant enough. Nevertheless, since the correlations between the $K_{\rm{sym}}$ and $J_{\rm{sym}}$ are constrained along the boundaries, the uncertainty due to $J_{\rm{sym}}$ is not as big as if it is allowed to vary freely between -200 and 800 MeV. The above observational boundaries will be used to set limits on the high-density EOS and symmetry energy as we shall discuss in detail next.

\section{Combined astrophysical constraints on the pressure in neutron star matter} \label{com1}
\begin{figure}
\begin{center}
\resizebox{0.55\textwidth}{!}{
  \includegraphics{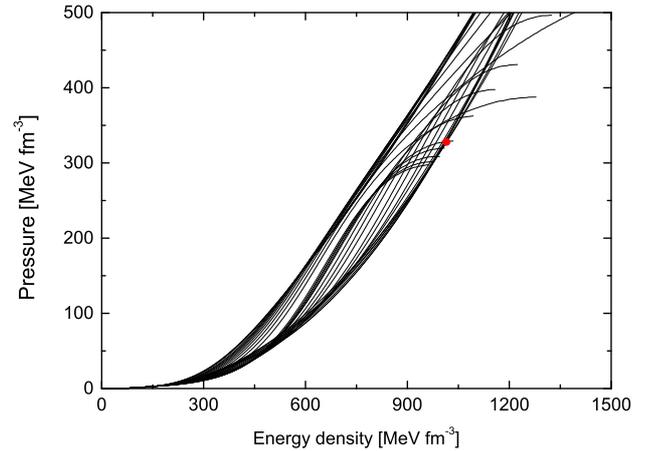}
}
\end{center}
  \caption{(color online) Examples of pressure in neutron stars at $\beta$ equilibrium satisfying all constraints considered in this work. A similar plot can be found in ref. \cite{NBZ-NST}.}
\label{EOS-exams}       
\end{figure}
Within the allowed region in the $K_{\rm{sym}}$-$J_{\rm{sym}}$ plane, the $J_0$ is limited from above by the causality surface and below by the mass constraint M$_{\rm{max}}\geq 2.01$ M$_{\odot}$, respectively.  The
pressure of NS matter at $\beta$ equilibrium in this allowed 3D space can be calculated by using the formalism given in Sec. \ref{tov}. 

\begin{figure*}
\begin{center}
\resizebox{0.46\textwidth}{!}{
  \includegraphics{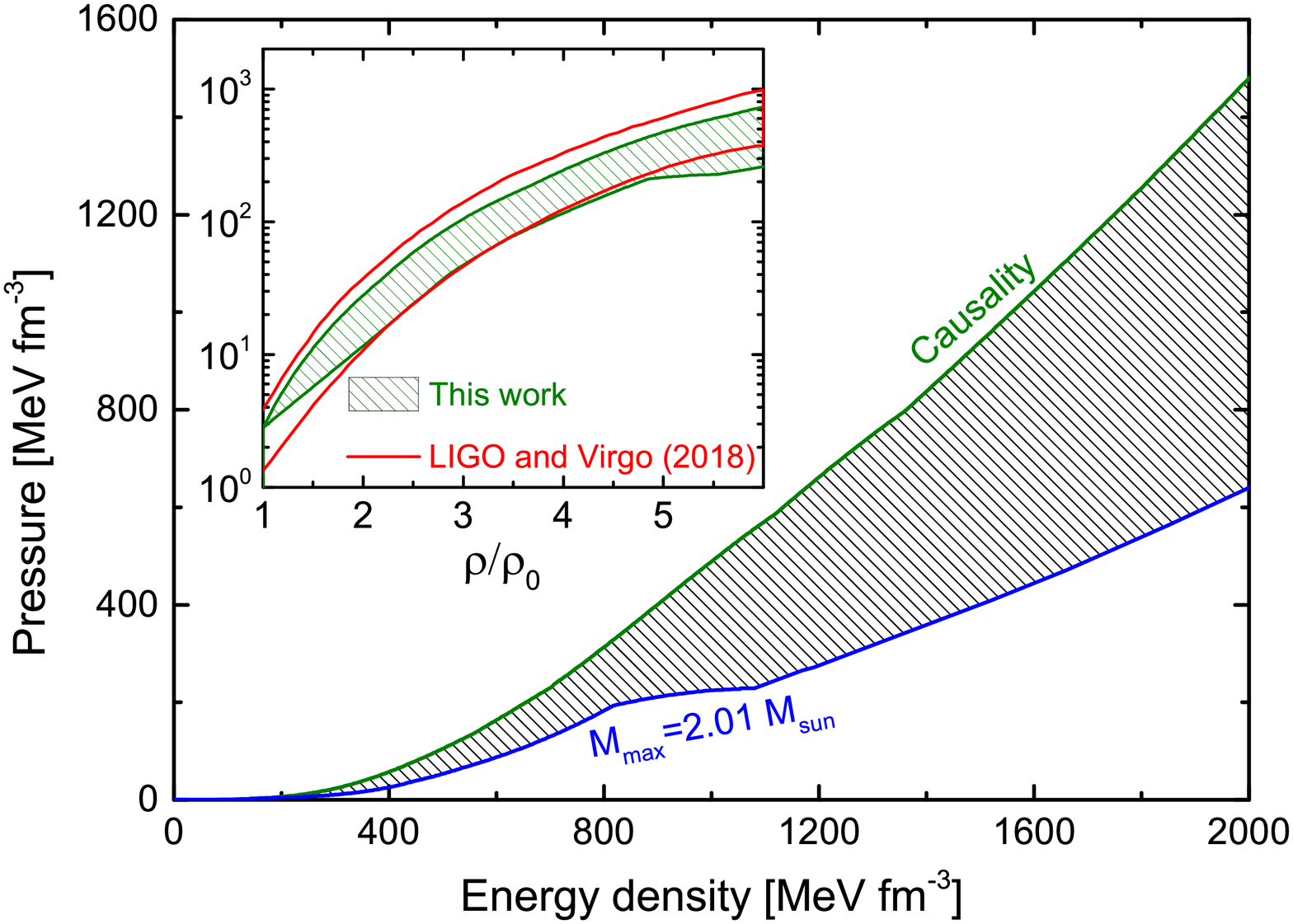}
  }
  \resizebox{0.46\textwidth}{!}{
  \includegraphics{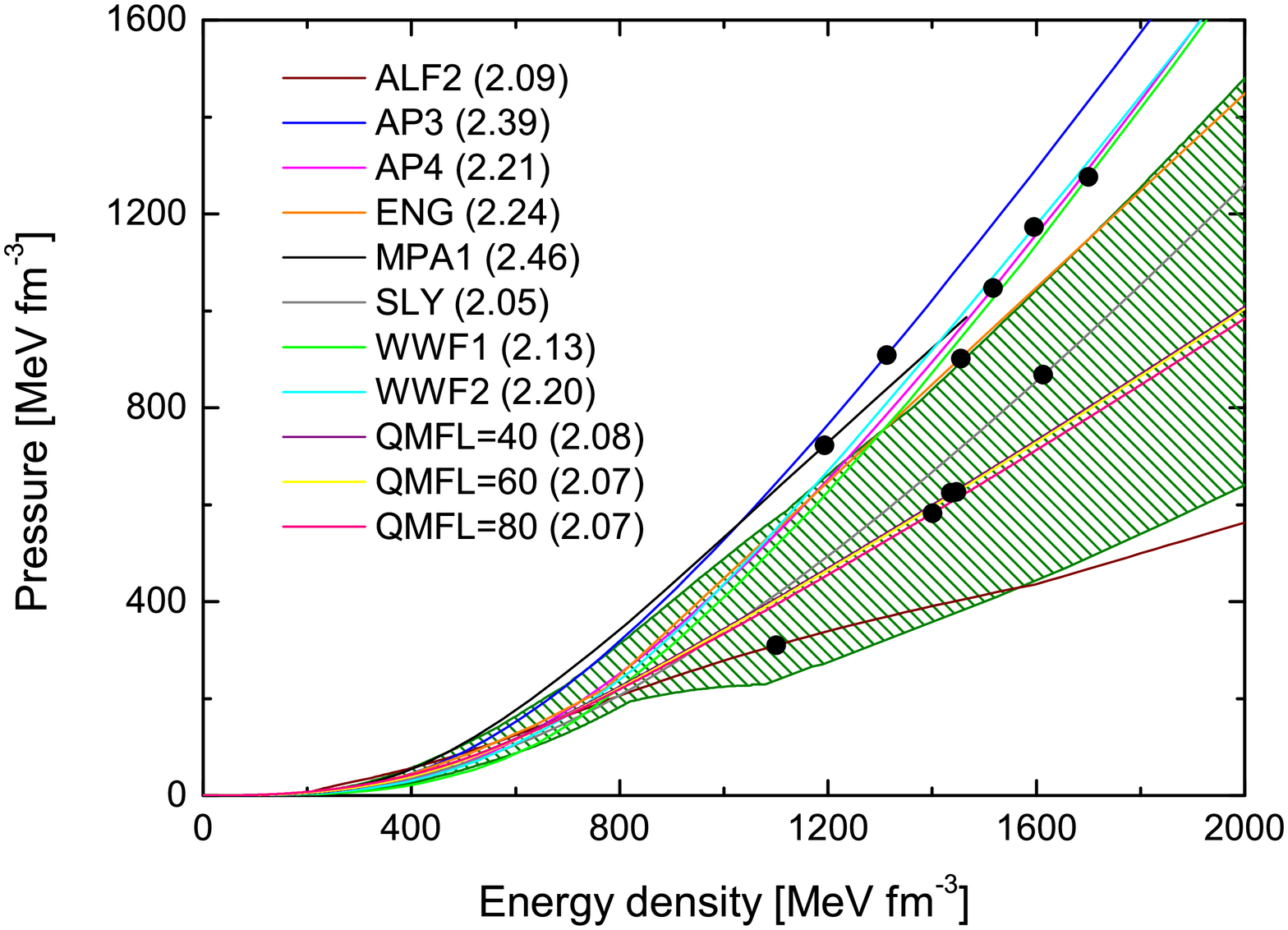}
}
\end{center}
  \caption{(color online) Left: the pressure as a function of energy density (baryon density in the inset) in neutron star matter at $\beta$ equilibrium extracted in this work from the astrophysical observational constraints (shaded regions). A comparison with the LIGO+Virgo result at 90\% confidence level (red boundary) \cite{LIGO2018} is shown in the inset. Right: Comparisons of the extracted pressure in this work with the predictions of several EOS models predicting maximum masses higher than 2.01 M$_{\odot}$. The maximum masses predicted are in parentheses following the model names. The black symbols indicate the maximum pressure and energy density reached at the maximum mass predicted in the specified model.
 }
\label{observations}       
\end{figure*}

\subsection{How we extract the upper and lower limit of nuclear pressure using observations of neutron stars and the causality condition}
Shown in Fig. \ref{EOS-exams} are examples of NS pressure as a function of
energy density. The lower and upper boundaries of the pressure satisfying all constraints discussed above are extracted by finding the lowest/highest pressure at a given energy density.
As expected, as the three high-density EOS parameters $K_{\rm{sym}}$, $J_{\rm{sym}}$ and $J_0$ vary, some EOSs keep increasing while some others have already increased to their maximum energy densities allowed by the
constraints in the energy density range plotted. Thus, in determining the lowest pressure at a fixed energy density, some softening along the lower boundary of the pressure, e.g., around the red point, may appear due to the crossing of the EOSs with different growing tendencies.
The constraining band on the pressure in NSs at $\beta$ equilibrium extracted using the approach described above is shown as a function of energy density in the two windows of
Fig. \ref{observations}.  The softening of the lower boundary around the energy density range of 800 to 1000 MeV/fm$^3$ is due to the EOS crossings discussed above.
While the symmetry energy contributes to the pressure and determines the composition of NSs at $\beta$ equilibrium, the total pressure calculated self-consistently is dominated
by the SNM contribution at high densities. The upper and lower boundaries of the pressure are thus dominated by the causality and mass constraint M$_{\rm{max}}\geq 2.01$ M$_{\odot}$ through the $J_0$ parameter,
while the variation of the spread in pressure is due to the variation of all three high-density EOS parameters.

\begin{figure*}
\begin{center}
\resizebox{0.45\textwidth}{!}{
  \includegraphics{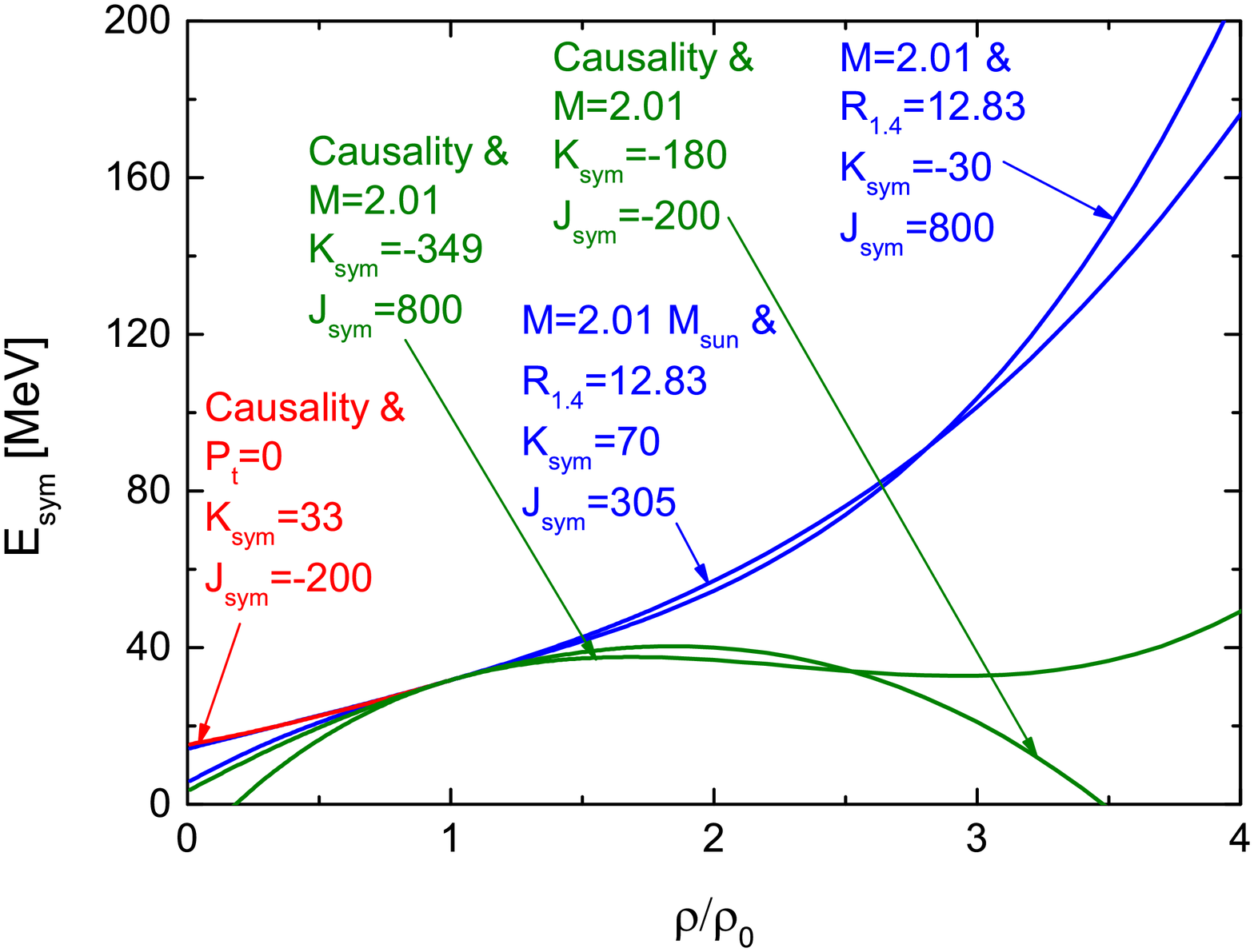}
  }
  \resizebox{0.45\textwidth}{!}{
 \includegraphics{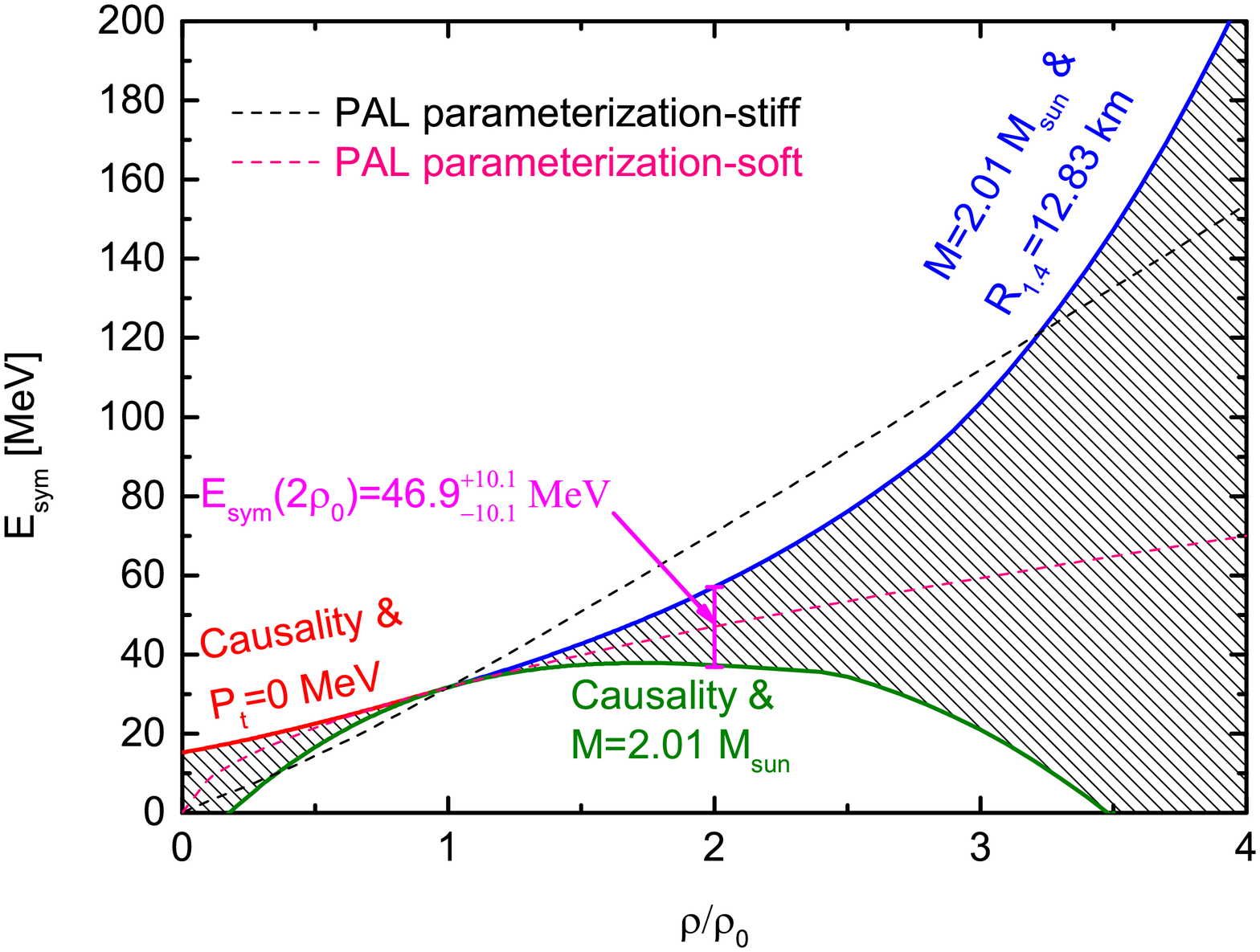}
}
\caption{Left: demonstrations of how each segment of the upper and lower boundaries of the symmetry energy are obtained using the boundaries in the $K_{\rm{sym}}$-$J_{\rm{sym}}$
plane shown in Fig. \ref{exc1}. Right: a comparison with the Prakash, Ainsworth and Lattimer parameterizations of nuclear symmetry energy.}
\label{Esym-plot}
\end{center}
\end{figure*}

\subsection{Nuclear pressure in neutron stars extracted using their radii in comparison with that from the tidal deformability in GW170817 and predictions using microscopic nuclear many-body theories}
The inset on the left in Fig. \ref{observations} is the pressure as a function of baryon density in comparison with the pressures at 90\% confidence level extracted by the LIGO+Virgo Collaborations.
In our approach, the upper and lower boundaries represent 100\% confidence level of finding the pressure to be between them. With this understanding, the overlap of the two results is overwhelming.
Since we did not use the tidal deformability in deriving the constraining band on pressure, there is no self-correlation in the comparison with the LIGO+Virgo result.
On the other hand, it is also not surprising since a number of independent analyses have extracted from the tidal deformability of GW170817 the upper limit on the radius $R_{1.4}$ consistent with the value
$R_{1.4}\leq 12.83$ km we used in deriving the pressure.

It is also useful to compare the extracted pressure band with predictions using EOSs available in the literature in the right window of Fig. \ref{observations}. We only selected typical EOSs that can support a maximum mass of NSs higher than 2.01 M$_{\odot}$. The EOSs are ALF2 of Alford et al. \cite{ALF2} for hybrid (nuclear + quark matter) stars,  APR3 and APR4 of Akmal and Pandharipande \cite{AP34}, ENG of Engvik et al. \cite{ENG}, MPA1 of Muther, Prakash and Ainsworth \cite{MPA1}, SLy of Douchin and Haensel \cite{SLy}, WWF1 and WWF2 of Wiringa, Fiks and Fabrocini \cite{WFF12}, the QMFL40, QMFL60 and QMFL80 within the Quark Mean Field model with L=40, 60 and 80 MeV, respectively, from the recent work of Zhu et al. \cite{QMF}. The individual predictions of the maximum masses are in parentheses following the model names in Fig. \ref{observations}. The black symbols indicate the maximum pressure and energy density reached at the maximum mass supported by the EOS of the mode. While the constraining band for the pressure is still quite wide especially at high energy densities,
predictions of several of these theoretical EOSs are too stiff to be bounded by the constraining band from our analyses of the astrophysical observations.

\section{Combined astrophysical constraints on nuclear symmetry energies at high densities}\label{com2}
We now turn to the combined astrophysical constraints on nuclear symmetry energies at high densities. This is achieved by examining the $E_{\rm{sym}}(\rho)$ functions using limiting $K_{\rm{sym}}$ and $J_{\rm{sym}}$
parameters on the constraining boundaries shown in Fig. \ref{exc1}. The shaded area is allowed by all astrophysical observables considered. While the $K_{\rm{sym}}$ is limited from left and right by the astrophysical observations, the
$J_{\rm{sym}}$ is still not limited. We therefore will consider the whole range of $J_{\rm{sym}}$ from -200 to 800 along the boundaries in extracting the astrophysical constraining band for $E_{\rm{sym}}(\rho)$. Fortunately, since the
$J_{\rm{sym}}$ controls the contribution of the $(\frac{\rho-\rho_0}{3\rho_0})^3$ term and its value is correlated to that of $K_{\rm{sym}}$ along the boundaries,
the extracted $E_{\rm{sym}}(\rho)$ is almost not affected at all by the large uncertainty of $J_{\rm{sym}}$ below about $2.5\rho_0$. At higher densities, it becomes gradually more important.
In the sub-saturation density region, the crossline between the causality surface and the thermodynamical stability condition (indicated as crust-core transition pressure $P_t=0$) as well as that between the causality surface and
the M$_{\rm{max}}\geq 2.01$ M$_{\odot}$ condition sets respectively the upper and lower boundaries for the $E_{\rm{sym}}(\rho)$. However, these limits are not practically useful since all model predictions fall between these limits in the sub-saturation density region.

\subsection{Extracting new constraints on nuclear symmetry energy around twice the saturation density of nuclear matter from astrophysical observations}\label{tworho}
The upper limit of $E_{\rm{sym}}(\rho)$ is determined by the left boundary in the $K_{\rm{sym}}-J_{\rm{sym}}$ plane shiwn in Fig. \ref{exc1}. There are three sections along this boundary as we discussed earlier. For the section describing the crossline between the M$_{\rm{max}}=2.01$ M$_{\odot}$ and $R_{1.4}=12.83$ km surfaces, its two ends on the left and right have the coordinates ($K_{\rm{sym}}=70$, $J_{\rm{sym}}=305$ MeV)
and ($K_{\rm{sym}}=-30$, $J_{\rm{sym}}=800$ MeV), respectively. The symmetry energy $E_{\rm{sym}}(\rho)$ with these two limiting parameter sets are shown with the two blue lines in the left window of Fig. \ref{Esym-plot}. It is seen that they are almost identical for $\rho_0\leq \rho \leq 3\rho_0$. Appreciable differences appear gradually at higher densities. To set conservative boundaries for $E_{\rm{sym}}(\rho)$, we choose the higher value as the upper limit. Similarly, for setting a conservative lower limit we compare the symmetry energy functions using two right most coordinates of the right boundary in the $K_{\rm{sym}}-J_{\rm{sym}}$ plane shown in Fig. \ref{exc1} at
($K_{\rm{sym}}=-180$, $J_{\rm{sym}}=-200$ MeV)
and ($K_{\rm{sym}}=-349$, $J_{\rm{sym}}=800$ MeV), respectively.  They lead to almost identical $E_{\rm{sym}}(\rho)$ for $\rho_0\leq \rho \leq 2.5\rho_0$. At higher densities, we choose the smaller one as the lower limit.

Shown in the right window of Fig. \ref{Esym-plot} are the extracted conservative bounds on the $E_{\rm{sym}}(\rho)$.
Certainly there are uncertainties associated with the boundaries we extracted here mainly because of the large remaining uncertainty of the $J_{\rm{sym}}$ parameter. Since the currently measured lower limit of the NS radius is not restrictive compared to the crossline between the causality surface and the M$_{\rm{max}}\geq 2.01$ M$_{\odot}$ condition, to reduce the gap between the upper and lower boundaries of
the $E_{\rm{sym}}(\rho)$ requires more precise measurements of NS radii or the measurement of at least another independent observable. Nevertheless, it is seen that both the upper and lower limits below about $2.5\rho_0$ are rather robust without being affected by the uncertainties from $J_{\rm{sym}}$. In fact, a reasonably tight constraint of $E_{\rm{sym}}(2\rho_0)=46.9\pm10.1$ MeV is obtained for the symmetry energy at twice the saturation density. The associated 21\% uncertainty is about twice the uncertainty of the most probable symmetry energy at saturation density $E_{\rm{sym}}(\rho_0)=31.7\pm 3.2$ MeV from combing 53 analyses of different kinds of experiments and observations accumulated over the last two decades.  On the other hand,  the slope of the symmetry energy $L_2\equiv 3\rho_0[\partial E_{\rm{sym}}(\rho)/\partial\rho]|_{\rho=2\rho_0}$ at $2\rho_0$ is about $L_2(2\rho_0)=56\pm 65$ MeV.  The large uncertainty of $L_2(2\rho_0)$ characterizes the spread of the symmetry energy at higher densities.

\subsection{Comparisons with existing constraints from terrestrial experiments and empirical parameterizations}\label{comparison}
As a reference, it is useful to compare the above values for $E_{\rm{sym}}(2\rho_0)$ and $L_2(2\rho_0)$ with extrapolations by Lie-Wen Chen using the systematics of $E_{\rm{sym}}(\rho)$ with its parameters constrained at and below the saturation density \cite{LWC15}. Using correlations of low-density $E_{\rm{sym}}(\rho)$ parameters supported by calculations of microscopic many-body theories,
and the reliable knowledge of $E_{\rm{sym}}({\rho_{0}}) = 32.5 \pm 0.5$ MeV, $E_{\rm{sym}}({\rho_c}) = 26.65 \pm 0.2$ MeV
and $L({\rho_c}) = 46.0 \pm 4.5$ MeV at a so-called cross density $\rho_{\rm{c}}= 0.11$ fm$^{-3}$ found from studying
nuclear masses and the neutron skin thicknesses of Sn isotopes, Chen extrapolated the systematics to supra-saturation densities and found that $E_{\rm{sym}}({2\rho _{0}}) \approx 40.2 \pm 12.8$
MeV and $L({2\rho _{0}}) \approx 8.9 \pm 108.7$ MeV. Interestingly, his value for $E_{\rm{sym}}({2\rho _{0}})$ is close to what we extracted here while his $L({2\rho _{0}})$ is much smaller and with an even larger uncertainty than what we have found here from studying the NS properties.

In a number of astrophysics \cite{Lattimer01} and heavy-ion reaction \cite{Li08} studies, the PAL (Prakash, Ainsworth and Lattimer) parameterizations \cite{Prakash88}
\begin{eqnarray}
E_{\rm{sym}}(\rho)(\rm{stiff})&=&12.7(\rho/\rho_0)^{2/3}+38(\rho/\rho_0)^2/(1+\rho/\rho_0),\nonumber\\
E_{\rm{sym}}(\rho)(\rm{soft})&=&12.7(\rho/\rho_0)^{2/3}+19(\rho/\rho_0)^{1/2}\nonumber
\end{eqnarray}
are often used as examples of the ``stiff" and ``soft" nuclear symmetry energies. A recent comparison \cite{Tsang2018} of the pressure in NS matter calculated using constraints from
kaon production and nucleon flow in heavy-ion reactions with that from the LIGO+Virgo Collaborations \cite{LIGO2018} indicates that one can not clearly distinguish the two PAL parameterizations
in the density range of $1-3\rho_0$. At higher densities, there are some weak indications that the stiff symmetry energy is preferred. However, in both the analyses of ref. \cite{Tsang2018} and
ref. \cite{LIGO2018} polytropes knowing nothing about the underlying $E_{\rm{sym}}(\rho)$ are used to parameterize directly the pressure in the core. As we have shown in Fig. \ref{observations},
the pressure extracted from our analyses also agree well with that from the LIGO+Virgo Collaborations. We found that the upper/lower bound on the pressure is mainly determined by the uncertainty of the SNM EOS through the
$J_0$ parameter. Since the $E_{\rm{sym}}(\rho)$ has much less effects on the pressure itself at supra-saturation densities, it is thus hard to extract any reliable information about the high-density $E_{\rm{sym}}(\rho)$ from directly comparing the pressures. 

In fact, it was shown numerically in Fig. 146 of ref. \cite{Li08} using the pressure of $npe$ matter in Eq. \ref{pre-npe} that the isospin-asymmetric pressure $P_{asy}$ in NSs at $\beta$ equilibrium dominates only near the saturation density. Above a transition density around $\rho_{\rm{transition}}=1.3\rho_0-2.5\rho_0$ depending on the stiffness of the symmetry energy, the total pressure is clearly and increasingly determined by the contribution from symmetric nuclear matter. Thus, there is a better chance to learn something about the $E_{\rm{sym}}(\rho)$ at densities below the $\rho_{\rm{transition}}$ from directly comparing the pressures inferred from heavy-ion reactions at intermediate energies and gravitational waves \cite{Tsang2018}. Our explicitly isospin-dependent parameterizations of the EOS at a more basic level through the specific nucleon energies in neutron-rich matter enabled us to extract more accurately and self-consistently the whole $E_{\rm{sym}}(\rho)$ itself underlying the pressure as a function of density. A comparison of the PAL parameterizations with the $E_{\rm{sym}}(\rho)$ constraining band in the right window of Fig. \ref{observations} clearly favors PAL's $E_{\rm{sym}}(\rho)(\rm{soft})$ parameterization.

Finally but not the least, it is also interesting to briefly compare with existing constraints on the symmetry energy around twice the saturation density from heavy-ion reactions with beam energies around 0.4-1.5 GeV/nucleon \cite{Trau18}. While the astrophysical constraint extracted above does not rule out the tendency of decreasing symmetry energy at supra-saturation densities, the value of $E_{\rm{sym}}(2\rho_0)=46.9\pm10.1$ MeV is very close to the one extracted by the ASY-EOS collaboration from studying the relative/differential flow and/or yield ratios of nucleons and light clusters in heavy-ion reactions \cite{russ11,ASY-EOS}. It is significantly above the value extracted from an earlier analysis of the ratio of charged pions \cite{XiaoPRL}. Since the isovector potentials are much weaker than the isoscalar potentials in heavy-ion reactions with limited isospin asymmetries reached, it is vey difficult to extract accurately high-density symmetry energies from heavy-ion reactions \cite{Li17}, albeit not as challenging as observing and analyzing gravitational waves. It is broadly recognized that constraining nuclear symmetry energies at supra-saturation densities more tightly using heavy-ion reactions requires improved model analyses and more data. Fortunately, both are expected to come soon thanks to the ongoing better coordinated theoretical efforts \cite{Transport} and the new data taken at the advanced rare isotope beam facilities \cite{Jer18,Lynch18}. In short, terrestrial laboratory experiments and astrophysical observations both contribute to the joint multi-messenger approach for constraining the EOS of dense neutron-rich nuclear matter.

\begin{figure}
\begin{center}
\resizebox{0.45\textwidth}{!}{
  \includegraphics{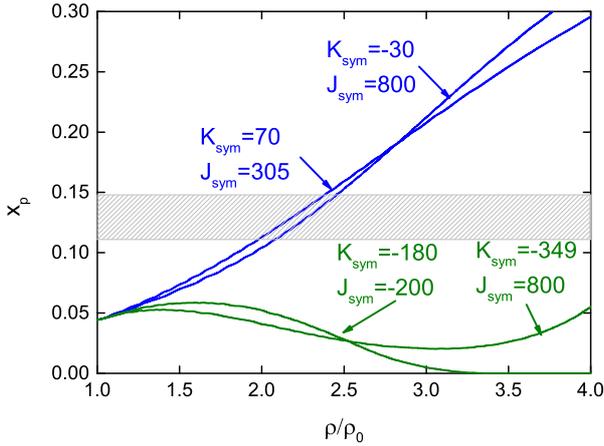}
  }
 \caption{Fractions of protons in neutron stars at $\beta$ equilibrium. The upper and lower boundaries correspond to those of the symmetry energy shown in the left window of 
Fig. \ref{Esym-plot}. The shaded band represents the minimum proton fraction between 11.1\% and 14/8\% for the direct URCA process to happen in the $npe\mu$ matter with the electron fraction varying between 1 (lower edge) and 0.5 (upper edge).}
\label{xprotons}
\end{center}
\end{figure}

\begin{figure}
\begin{center}
\resizebox{0.45\textwidth}{!}{
  \includegraphics{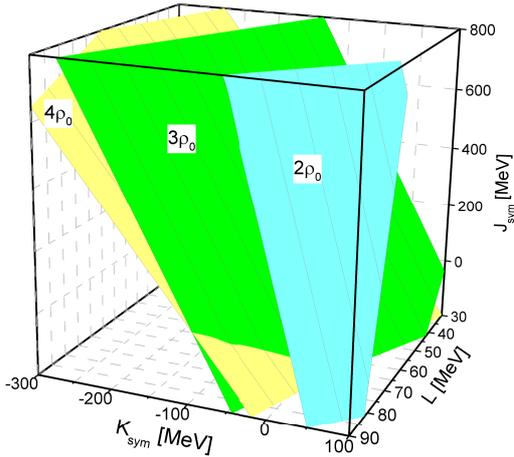}
  }
 \caption{Surfaces of constant critical densities for the direct URCA process to occur in the 3D symmetry energy parameter space. }
\label{DA}
\end{center}
\end{figure}

\subsection{Constraints on the proton fraction and critical density for direct URCA process}
An important reason for us to focus on the symmetry energy rather than the energy of PNM is that the
symmetry energy will give us information on composition of NSs.  The ranges of symmetry energies in Fig. \ref{Esym-plot} is staggering, and must give rise to a
large variation in compositions. It would thus be interesting to see the corresponding range for proton fractions. The results will help quantify the accuracy of models assuming NSs are made of only PNM. 
As we discussed earlier, the proton fraction $x_p$ at $\beta$ equilibrium at a given density is uniquely determined by the symmetry energy, if the quadratic approximation in Eq. (\ref{EOS0}) for 
the EOS of ANM is approximately correct. Shown in Fig. \ref{xprotons} are fractions of protons at supra-saturation densities in neutron stars at $\beta$ equilibrium. The upper and lower boundaries correspond to those of the symmetry energy shown in the left window of Fig. \ref{Esym-plot}. It is seen that the upper boundary runs between about 5\% to 30\% in the density range of $(1-4)\rho_0$. At twice the saturation density, the proton fraction is restricted within about (4-11)\% while at higher densities it can be as high as about 30\%. 

Depending on the physical quantities and the density ranges considered, the extracted proton fraction may or may not be important. For example, when one considers the energy per nucleon in ANM and the pressure in NSs near the saturation density $\rho_0$, the small proton fraction of about 5\% indicates that the PNM approximation is almost perfect. But it does not mean that the information about the symmetry energy is not important. In fact, the pressure in PNM at saturation density is completely determined by the slope parameter L of the symmetry energy within the parabolic approximation of the ANM EOS. Even at twice the saturation density, the energy in NS matter may be well approximated by that of PNM. While the pressure there has significant contributions from both the SNM and the symmetry energy related terms (see, e.g., Eq. (\ref{pre-npe}) for the pne matter). At higher densities, although the proton fraction is at most 30\% in the density range considered, the symmetry energy still plays significant roles. Comparing the results in Fig. \ref{observations} and Fig. \ref{Esym-plot}, it is seen that around twice the saturation density the spread in pressure is about a factor of 2.5 while that in symmetry energy is only about 20\%. As we discussed earlier, the spread in pressure is mainly due to the uncertainty in the $J_0$ parameter describing the high-density EOS of SNM. Again, this indicates the challenges of inferring the symmetry energy directly from the pressure without using an explicitly isospin-dependent EOS unless the exact information about the proton fraction is known already. Unfortunately, such information is not available {\it a priori}. 

As discussed in the seminal work of Lattimer, Pethick, Prakash and Haensel in ref. \cite{LPPH} and many subsequent studies, the density dependence of nuclear symmetry energy and the resulting proton fraction are most critical for determining the cooling mechanisms of protoneutron stars. In the $npe\mu$ matter the threshold proton fraction $x^{DU}_p$ necessary for the fast cooling through the so-called direct URCA process (DU) to occur
\begin{equation}\label{xdu}
x^{DU}_p=1/[1+(1+x_e^{1/3})^3]
\end{equation} 
is between 11.1\% to 14.8\% for the electron fraction $x_e\equiv \rho_e/\rho_p$ between 1 and 0.5 \cite{Klahn}.  This range is indicated by the horizontal band in Fig. \ref{xprotons}. It is seen that even along the upper boundary of the 
allowed proton fraction, the direct URCA will require densities higher than about twice the saturation density. For softer symmetry energies, higher critical densities are required for the direct URCA cooling mechanism to function. 

To further illustrate the roles of the three parameters (L, $K_{\rm sym}$ and $J_{\rm sym}$) describing the density dependence of nuclear symmetry energy in different density regions, we show in Fig. \ref{DA} surfaces of constant critical densities $\rho^{DU}_c$ for the direct URCA process to occur in the 3D symmetry energy parameter space. The critical density is determined by requiring $x_p=x^{DU}_p$ with the electron fraction $x_e$ in the $npe\mu$ matter calculated consistently from the charge neutrality condition in solving the TOV equation. To construct the 3D surfaces of constant $\rho^{DU}_c$ values, we find all combinations of the symmetry energy parameters giving the same specified value of each $\rho^{DU}_c$ based on Eq. (\ref{xdu}). We start from setting $\rho^{DU}_c=2\rho_0$ as our results in Fig. \ref{xprotons} have already indicated that the upper boundary for the proton fraction allows the direct URCA to happen above $2\rho_0$. If the proton fraction is below the upper bound, then even high critical densities are required. It is seen that the $\rho^{DU}_c=2\rho_0$ constant surface is rather vertical and narrow in $L$ indicating that the high-density symmetry energy parameters $K_{\rm sym}$ and $J_{\rm sym}$ are not so important for determining this relatively low density $\rho^{DU}_c$ as one can easily understand, while the allowed range of $L$ is small because around and below $2\rho_0$ the symmetry energy has been relatively well constrained as we discussed earlier. By setting the $\rho^{DU}_c$ at higher densities, however, the $K_{\rm sym}$ and $J_{\rm sym}$ start to play more significant roles as one expects. For example, for $\rho^{DU}_c=3\rho_0$ broad ranges of all three parameters are allowed. In particular, when the $J_{\rm sym}$ is low, the $K_{\rm sym}$ needs to be high and the $L$ can have any value between 30 and 90 MeV. On the other hand, when the $J_{\rm sym}$ is high, the $K_{\rm sym}$ needs to be low but the $L$ has little effect. These observations are consistent with the effects of varying the three parameters on the density dependence of nuclear symmetry energy as we demonstrated earlier in Fig. \ref{Ksymeffect}. It is also interesting to note that for the super-soft symmetry energies, i.e., those that first increase then decrease at high densities with negative values of $K_{\rm sym}$ and/or $J_{\rm sym}$,  the equation $x_p=x^{DU}_p$ has two physically meaningful roots. The direct URCA process can happen between the two critical densities. This results in the crossing of two constant surfaces with large values of $\rho^{DU}_c$, such as those with $\rho^{DU}_c=3\rho_0$ and $\rho^{DU}_c=4\rho_0$, in regions where the $K_{\rm sym}$ and/or $J_{\rm sym}$ are very low, leading to super-soft symmetry energies.

Compared to the constraints on the symmetry energy from astrophysical observations on the radii, mass and tidal deformability, the cooling of neutron stars can in principle directly constrain the high-density behavior of nuclear symmetry energy. Of course, the cooling of NSs involves many interesting but still uncertain physics and the cooling data are also limited. Nevertheless, we are very optimistic that new observations providing evidences of direct URCA processes may give us additional and hopefully direct information about nuclear symmetry energies at high densities \cite{EBrown}. 
\\

\subsection{Impacts on nuclear many-both theory predictions and limitations of the high-density symmetry energy extracted here using astrophysical observations}
The strong power of our extracted constraining band on the $E_{\rm{sym}}(\rho)$ in distinguishing existing model predictions is clearly demonstrated in Fig. \ref{Esym-survey}.
Although the band is still quite wide above about $2.5\rho_0$, like it or not, a large number of predictions especially those based on phenomenological models or energy density functionals run out the constraining boundaries.
Compared to the spread of predicted $E_{\rm{sym}}(2\rho_0)$ values between $15$ MeV to $100$ MeV by various theories shown in  Fig. \ref{Esym-survey}, the value of $E_{\rm{sym}}(2\rho_0)=46.9\pm10.1$ MeV extracted here represents a significant progress in the field.

It is worth emphasizing that in our current study the most probable values of the empirical saturation properties are used as we mentioned earlier. All of them still have some uncertainties. Since the lower bound we extracted is from the crossline of causality and the maximum mass condition M$_{\rm{max}}\geq 2.01$ M$_{\odot}$ and none of them depends sensitively on the saturation properties, the lower bound is expected to be approximately the same if we loose the constraints on the saturation properties. On the other hand, since the upper bound is from the conditions of M$_{\rm{max}}\geq 2.01$ M$_{\odot}$ and $R_{1.4}\leq 12.83$ km and the radius is known to have some dependences on the $L$ parameter, if we loose the constraint on the later, the upper bound is expected to be altered. While we expect this possible modification is small and our bounds are already rather conservative, an investigation of the probability distribution functions of all the $E_{\rm{sym}}(\rho)$ parameters within the Bayesian framework is underway. Effects of all uncertainties and their correlations will be studied and reported elsewhere. The upper/lower bound we extracted here from basically only two NS observables represents the most probable bound under the conservative conditions used.

\section{Summary and outlook}
In summary, the density dependence of nuclear symmetry energy $E_{\rm{sym}}(\rho)$ is the most uncertain part of the EOS of neutron-rich nucleonic matter especially at supra-saturation densities. Essentially all available nuclear many-body theories and interactions have been used to predict the $E_{\rm{sym}}(\rho)$. However, the predictions diverge widely especially at high densities. Among the difficulties of predicting accurately the $E_{\rm{sym}}(\rho)$ are  our poor knowledge about the weak isospin-dependence of strong force, the spin-isospin dependence of three-body nuclear forces and the tensor-force induced isospin-dependence of short-range nucleon-nucleon correlations in dense matter besides the challenges of solving accurately nuclear many-body problems. It is well known that the $E_{\rm{sym}}(\rho)$ has many important effects on properties of neutron-rich nuclei and nuclear reactions. It also affects some properties of NSs and gravitational waves from sources/events involving NSs.

Using an explicitly isospin-dependent parametric EOS with three parameters characterizing nucleon specific energy in dense neutron-rich matter, for a single given NS observable, such as its mass, radius or tidal deformability we can find all required combinations of the EOS parameters by inverting numerically the TOV equation. Compared to the widely used isospin-independent polytropes of pressure as a function of energy or baryon density for the core of NSs, our isospin-dependent parameterization of nucleon specific energy is at a more basic level and necessary for extracting the underlying nuclear symmetry energy. Applying our approach using observational
constraints on the radii, maximum mass and tidal deformability of NSs as well as the causality condition all together, we have learned the following new and important physics compared to the existing knowledge in the literature:

\begin{itemize}
\item By studying the variation of causality surface where the speed of sound is the same as that of light at central densities of the most massive NSs within the uncertain ranges of high-density EOS parameters, the absolutely maximum mass of NSs is found to be 2.40 M$_{\odot}$ approximately independent of the EOSs used. This limiting mass is consistent with the findings of several recent analyses about the maximum mass of the possible super-massive remanent produced in the immediate aftermath of GW170817.
 \\
 \item Boundaries are established for the high-density EOS parameter space by examining the crosslines of the minimum maximum mass M$_{\rm{max}}\geq 2.01$ M$_{\odot}$, the radius range of $R_{\rm{1.4}}$= 10.62-12.83 km for
canonical NSs as well as the causality surface. These boundaries lead to constraining bands for both the pressure as a function of energy (baryon) density and the density dependence of nuclear symmetry energy.
Our EOS presented using pressure as a function of energy density is in good agreement with that extracted recently by the LIGO+Virgo Collaborations from their improved analyses of the NS tidal deformability in GW170817.
\\
\item The pressure constraining band is also compared with predictions of several typical EOSs available in the literature. Several predictions were found to run out of the constraining band at high densities.
The pressure boundaries are mostly determined by the minimum maximum mass and causality conditions with little influences from
variations of nuclear symmetry energy, making it difficult to extract any reliable information about the high-density nuclear symmetry energy from studying directly the total pressure of NS matter itself.
\\
\item Rather robust upper and lower boundaries for the {DU}symmetry energy are extracted up to about $2.5\rho_0$ while at higher densities the boundaries suffer some uncertainties. The upper bound is obtained from
the crosslines of the $R_{\rm{1.4}}=12.83$ km and M$_{\rm{max}}=2.01$ M$_{\odot}$ surfaces in the 3D EOS parameter space, while the lower one is from the crossline of the causality and M$_{\rm{max}}=2.01$ M$_{\odot}$
surfaces. Many available predictions for nuclear symmetry energy run out of the extracted boundaries at various densities.
The symmetry energy at $2\rho_0$ is constrained to $E_{\rm{sym}}(2\rho_0)=46.9\pm10.1$ MeV excluding many of the existing predictions scattered between $E_{\rm{sym}}(2\rho_0)=15$ and 100 MeV. Thus, the $E_{\rm{sym}}(\rho)$ at supra-saturation densities  extracted in this work from observations of NSs represent a significant progress in the field compared to the prior knowledge in the literature.

\end{itemize}

At densities higher than about twice the saturation density of nuclear matter, the symmetry energy is still not well constrained by the astrophysical observables and physics conditions used here. To narrow down the $E_{\rm{sym}}(\rho)$ above $2\rho_0$, independent measurements of other observables and/or improvements of the accuracy of radius measurements are necessary. In addition, ongoing efforts using high-energy heavy-ion reactions at several radioactive beam facilities may also help further constrain nuclear symmetry energies above twice the saturation density. We are hopeful that eventually the multi-messengers approach of combining probes in both astrophysical observations and terrestrial experiments will lead us to a narrow stripe of nuclear symmetry energies at high densities.

\section*{Acknowledgement}
We would like to thank Lie-Wen Chen, Bin Qi, S.Y. Wang and Jun Xu for helpful discussions. NBZ is supported in part by the China Scholarship Council. BAL acknowledges the U.S. Department of Energy, Office of Science, under Award Number DE-SC0013702, the CUSTIPEN (China-U.S. Theory Institute for Physics with Exotic Nuclei) under the US Department of Energy Grant No. DE-SC0009971 and the National Natural Science Foundation of China under Grant No. 11320101004.
%
%

\end{document}